\documentclass[12pt]{article}
\usepackage{epsfig}
\usepackage{amsmath}
\usepackage{hhline}
\usepackage{amssymb}
\usepackage[english]{babel}
\usepackage{graphicx,rotating}
\usepackage{textcomp}

\newcommand{\starpar}{\ensuremath{*)}}
\makeatletter
   \renewcommand{\@fnsymbol}[1]{%
    \ifcase#1\or\starpar\or \textdagger\or \textdaggerdbl
    \or\textsection\or \textparagraph\or \textbardbl
    \or \textasteriskcentered\textasteriskcentered 
    \or\textdagger\textdagger \or \textdaggerdbl\textdaggerdbl
    \else\@ctrerr\fi}
  \makeatother
\long\def\symbolfootnote[#1]#2{\begingroup
\def\thefootnote{\fnsymbol{footnote}}\footnote[#1]{#2}\endgroup}

\date{}
\newcommand{\updates}[1]%
        {\fbox{\parbox{\linewidth}{\textbf{Updates from last year:}\\#1}}}

\def\leqsim{\mathbin{\;\raise1pt\hbox{$<$}\kern-8pt\lower3pt\hbox{$\sim$}\;}}
\def\geqsim{\mathbin{\;\raise1pt\hbox{$>$}\kern-8pt\lower3pt\hbox{$\sim$}\;}}
\newcommand{\cls}{\ensuremath{{\rm CL}_{\rm s}}}
\newcommand{\ecls}{\ensuremath{\langle {\rm CL}_{\rm s} \rangle}}
\newcommand{\clb}{\ensuremath{{\rm CL}_{\rm b}}}

\newcommand{\ehad}{\ensuremath{E_{\rm had}}}
\newcommand{\mmiss}{\ensuremath{M_{\rm miss}}}
\newcommand{\nch}{\ensuremath{N_{\rm ch}}}
\newcommand{\msum}{\ensuremath{M_{\rm tot}}}
\newcommand{\msumf}{\ensuremath{M_{\rm tot}/\sqrt{s}}}
\newcommand{\mtot}{\ensuremath{M_{\rm tot}}}
\newcommand{\plsumf}{\ensuremath{P_{l}/\sqrt{s}}}
\newcommand{\ptsumf}{\ensuremath{P_{t}/\sqrt{s}}}
\newcommand{\ptsum}{\ensuremath{P_{t}}}
\newcommand{\pt}{\ensuremath{P_{t}}}

\newcommand{\esum}{\ensuremath{E_{\rm tot}}}
\newcommand{\emiss}{\ensuremath{E_{\rm miss}}}
\newcommand{\psum}{\ensuremath{\vec{P}_{\rm tot}}}
\newcommand{\pmiss}{\ensuremath{{P}_{\rm miss}}}
\newcommand{\yot}{\ensuremath{y_{12}}}

\newcommand{\ythf}{\ensuremath{y_{34}}}
\newcommand{\yffi}{\ensuremath{y_{45}}}
\newcommand{\yfis}{\ensuremath{y_{56}}}

\newcommand{\ffntmi}{\ensuremath{N^{\rm min}_{\rm ch,j4}}}
\newcommand{\fsntmi}{\ensuremath{N^{\rm min}_{\rm ch,j6}}}
\newcommand{\cthmiss}{\ensuremath{\cos(\theta_{\rm miss})}}
\newcommand{\ef}{\ensuremath{E_{12}}}

\newcommand{\fsanmi}{\ensuremath{\theta^{\rm min}_{\rm j6}}}
\newcommand{\fsanma}{\ensuremath{\theta^{\rm max}_{\rm j6}}}

\newcommand{\fsmo}{\ensuremath{M_{1}}}
\newcommand{\fsmt}{\ensuremath{M_{2}}}
\newcommand{\fsmth}{\ensuremath{M_{3}}}

\newcommand{\spher}{\ensuremath{S}}
\newcommand{\thrust}{\ensuremath{T}}
\newcommand{\ffanmi}{\ensuremath{\theta^{\rm min}_{\rm j4}}}
\newcommand{\ffanma}{\ensuremath{\theta^{\rm max}_{\rm j4}}}
\newcommand{\Ego}{\ensuremath{E_{\gamma_1}}}
\newcommand{\ELo}{\ensuremath{E_{\ell_1}}}
\newcommand{\ELt}{\ensuremath{E_{\ell_2}}}
\newcommand{\ELth}{\ensuremath{E_{\ell_3}}}

\newcommand{\PLo}{\ensuremath{\vec{P}_{\ell_1}}}
\newcommand{\PLt}{\ensuremath{\vec{P}_{\ell_2}}}
\newcommand{\PLth}{\ensuremath{\vec{P}_{\ell_3}}}

\newcommand{\ILo}{\ensuremath{I_{\ell_1}}}
\newcommand{\MZ}{\ensuremath{m_{\rm Z}}}
\newcommand{\MZrec}{\ensuremath{m_{\rm Z}^{\rm rec}}}
\newcommand{\TLoLt}{\ensuremath{\theta_{\ell_1 \ell_2}}}

\newcommand{\tLPtot}{\ensuremath{\theta_{\ell,\Sigma}}}
\newcommand{\MLPmiss}{\ensuremath{M_{\ell,P_{\rm miss}}}}
\newcommand{\nlacolin}{\ensuremath{\theta_{\rm aco\ (no\ lept)}}}
\newcommand{\nlthrust}{\ensuremath{T_{\rm no\ lept}}}
\newcommand{\nlyot}{\ensuremath{y_{\rm 12\ (no\ lept)}}}
\newcommand{\nlytth}{\ensuremath{y_{\rm 23\ (no\ lept)}}}
\newcommand{\nlythf}{\ensuremath{y_{\rm 34\ (no\ lept)}}}
\newcommand{\acolin}{\ensuremath{\theta_{\rm aco}}}
\newcommand{\aplan}{\ensuremath{A}}
\newcommand{\nlmsum}{\ensuremath{M_{\rm tot\ (hadr)}/\sqrt{s}}}
\newcommand{\nlmsumn}{\ensuremath{M_{\rm tot\ (hadr)}}}

\newcommand{\acopl}{\ensuremath{\Phi_{\rm aco}}}

\newcommand{\gaga}{\ensuremath{\gamma \gamma}}
\newcommand{\gev}{\ensuremath{{\rm\,GeV}}}
\newcommand{\gevc}{\ensuremath{{\rm\,GeV}/{ c}}}
\newcommand{\gevct}{\ensuremath{{\rm\,GeV}/{ c}^2}}

\def\NPB#1#2#3{{\rm Nucl.~Phys.} {\bf{B#1}} (#2) #3}
\def\PLB#1#2#3{{\rm Phys.~Lett.} {\bf{B#1}} (#2) #3}

\def\PRL#1#2#3{{\rm Phys.~Rev.~Lett.} {\bf{#1}} (#2) #3}

\def\EPJ#1#2#3{{\rm Eur.~Phys.~J.} {\bf C#1} (#2) #3}

\def\NIMA#1#2#3{{\rm Nucl.~Instrum.~and~Methods} {\bf{A#1}} (#2) #3}
\def\CPC#1#2#3{{\rm Comput.~Phys.~Commun.} {\bf#1} (#2) #3}

\newcommand{\sqrts}{\ensuremath{\sqrt{s}}}
\newcommand{\pbm}{\ensuremath{{\rm pb}^{-1}}}

\begin{document}
\font\eightrm=cmr8
\font\ninerm=cmr9

\begin{titlepage}

\vskip -3.0cm
\begin{center}
   {\large EUROPEAN ORGANIZATION FOR NUCLEAR RESEARCH (CERN)}
\end{center}

\begin{flushleft}
   \end{flushleft}
   \begin{flushright}
      CERN-PH-EP/2006-XXX\\
      May 11, 2006\\
\end{flushright}
\vskip 2.3cm

\begin{center}
   \boldmath
   \centerline{\LARGE\bf Search for Higgs bosons decaying to WW }
   \vskip .2 cm
   \centerline{\LARGE\bf in e$^+$e$^-$ collisions at LEP}
   \vskip 1. cm
   \unboldmath
   The ALEPH Collaboration \symbolfootnote[1]{See next pages for the list of authors}
\end{center}
\vskip 1.8cm

\begin{abstract}
\vspace{.2cm}
A search for Higgs bosons produced in association with a fermion pair, and decaying to $\mathrm{WW}$,
is performed with the data collected by the ALEPH detector at centre-of-mass energies ranging from 
191 to 209\gev. The data correspond to an integrated luminosity of 453.2 pb$^{-1}$. 
Thirteen exclusive selections are developed according to the different final state topologies. 
No statistically significant evidence for a Higgs boson decaying into a WW pair has been found.
An upper limit is derived, as a function of the Higgs boson mass, on the product of the e$^+$e$^-$ $\rightarrow$ $\mathrm{Hf\bar{f}}$ cross section and the $\mathrm{H \to WW}$ branching ratio.
The data on the search for $\mathrm{H \to WW}$ are combined with previously published ALEPH results on the search for $\mathrm{H \to \gamma\gamma}$, to significantly extend the limits on the mass of a fermiophobic Higgs boson.
\end{abstract}
\vskip 1.5cm
\begin{center}
Submitted to Eur. Phys. J.C
\end{center}
\vfill
\end{titlepage}
\pagestyle{empty}
\newpage
\small
%
%
\newlength{\saveparskip}
\newlength{\savetextheight}
\newlength{\savetopmargin}
\newlength{\savetextwidth}
\newlength{\saveoddsidemargin}
\newlength{\savetopsep}
\setlength{\saveparskip}{\parskip}
\setlength{\savetextheight}{\textheight}
\setlength{\savetopmargin}{\topmargin}
\setlength{\savetextwidth}{\textwidth}
\setlength{\saveoddsidemargin}{\oddsidemargin}
\setlength{\savetopsep}{\topsep}
%
%
\setlength{\parskip}{0.0cm}
\setlength{\textheight}{25.0cm}
\setlength{\topmargin}{-1.5cm}
\setlength{\textwidth}{16 cm}
\setlength{\oddsidemargin}{-0.0cm}
\setlength{\topsep}{1mm}
\pretolerance=10000
\centerline{\large\bf The ALEPH Collaboration}
\footnotesize
\vspace{0.5cm}
{\raggedbottom
\begin{sloppypar}
\samepage\noindent
S.~Schael,
\nopagebreak
\begin{center}
\parbox{15.5cm}{\sl\samepage
Physikalisches Institut das RWTH-Aachen, D-52056 Aachen, Germany}
\end{center}\end{sloppypar}
\vspace{2mm}
\begin{sloppypar}
\noindent
R.~Barate,
R.~Bruneli\`ere,
I.~De~Bonis,
D.~Decamp,
C.~Goy,
S.~J\'ez\'equel,
J.-P.~Lees,
F.~Martin,
E.~Merle,
\mbox{M.-N.~Minard},
B.~Pietrzyk,
B.~Trocm\'e
\nopagebreak
\begin{center}
\parbox{15.5cm}{\sl\samepage
Laboratoire de Physique des Particules (LAPP), IN$^{2}$P$^{3}$-CNRS,
F-74019 Annecy-le-Vieux Cedex, France}
\end{center}\end{sloppypar}
\vspace{2mm}
\begin{sloppypar}
\noindent
S.~Bravo,
M.P.~Casado,
M.~Chmeissani,
J.M.~Crespo,
E.~Fernandez,
M.~Fernandez-Bosman,
Ll.~Garrido,$^{15}$
M.~Martinez,
A.~Pacheco,
H.~Ruiz
\nopagebreak
\begin{center}
\parbox{15.5cm}{\sl\samepage
Institut de F\'{i}sica d'Altes Energies, Universitat Aut\`{o}noma
de Barcelona, E-08193 Bellaterra (Barcelona), Spain$^{7}$}
\end{center}\end{sloppypar}
\vspace{2mm}
\begin{sloppypar}
\noindent
A.~Colaleo,
D.~Creanza,
N.~De~Filippis,
M.~de~Palma,
G.~Iaselli,
G.~Maggi,
M.~Maggi,
S.~Nuzzo,
A.~Ranieri,
G.~Raso,$^{24}$
F.~Ruggieri,
G.~Selvaggi,
L.~Silvestris,
P.~Tempesta,
A.~Tricomi,$^{3}$
G.~Zito
\nopagebreak
\begin{center}
\parbox{15.5cm}{\sl\samepage
Dipartimento di Fisica, INFN Sezione di Bari, I-70126 Bari, Italy}
\end{center}\end{sloppypar}
\vspace{2mm}
\begin{sloppypar}
\noindent
X.~Huang,
J.~Lin,
Q. Ouyang,
T.~Wang,
Y.~Xie,
R.~Xu,
S.~Xue,
J.~Zhang,
L.~Zhang,
W.~Zhao
\nopagebreak
\begin{center}
\parbox{15.5cm}{\sl\samepage
Institute of High Energy Physics, Academia Sinica, Beijing, The People's
Republic of China$^{8}$}
\end{center}\end{sloppypar}
\vspace{2mm}
\begin{sloppypar}
\noindent
D.~Abbaneo,
T.~Barklow,$^{26}$
O.~Buchm\"uller,$^{26}$
M.~Cattaneo,
B.~Clerbaux,$^{23}$
H.~Drevermann,
R.W.~Forty,
M.~Frank,
F.~Gianotti,
J.B.~Hansen,
J.~Harvey,
D.E.~Hutchcroft,$^{30}$,
P.~Janot,
B.~Jost,
M.~Kado,$^{2}$
P.~Mato,
A.~Moutoussi,
F.~Ranjard,
L.~Rolandi,
D.~Schlatter,
F.~Teubert,
A.~Valassi,
I.~Videau
\nopagebreak
\begin{center}
\parbox{15.5cm}{\sl\samepage
European Laboratory for Particle Physics (CERN), CH-1211 Geneva 23,
Switzerland}
\end{center}\end{sloppypar}
\vspace{2mm}
\begin{sloppypar}
\noindent
F.~Badaud,
S.~Dessagne,
A.~Falvard,$^{20}$
D.~Fayolle,
P.~Gay,
J.~Jousset,
B.~Michel,
S.~Monteil,
D.~Pallin,
J.M.~Pascolo,
P.~Perret
\nopagebreak
\begin{center}
\parbox{15.5cm}{\sl\samepage
Laboratoire de Physique Corpusculaire, Universit\'e Blaise Pascal,
IN$^{2}$P$^{3}$-CNRS, Clermont-Ferrand, F-63177 Aubi\`{e}re, France}
\end{center}\end{sloppypar}
\vspace{2mm}
\begin{sloppypar}
\noindent
J.D.~Hansen,
J.R.~Hansen,
P.H.~Hansen,
A.C.~Kraan,
B.S.~Nilsson
\nopagebreak
\begin{center}
\parbox{15.5cm}{\sl\samepage
Niels Bohr Institute, 2100 Copenhagen, DK-Denmark$^{9}$}
\end{center}\end{sloppypar}
\vspace{2mm}
\begin{sloppypar}
\noindent
A.~Kyriakis,
C.~Markou,
E.~Simopoulou,
A.~Vayaki,
K.~Zachariadou
\nopagebreak
\begin{center}
\parbox{15.5cm}{\sl\samepage
Nuclear Research Center Demokritos (NRCD), GR-15310 Attiki, Greece}
\end{center}\end{sloppypar}
\vspace{2mm}
\begin{sloppypar}
\noindent
A.~Blondel,$^{12}$
\mbox{J.-C.~Brient},
F.~Machefert,
A.~Roug\'{e},
H.~Videau
\nopagebreak
\begin{center}
\parbox{15.5cm}{\sl\samepage
Laoratoire Leprince-Ringuet, Ecole
Polytechnique, IN$^{2}$P$^{3}$-CNRS, \mbox{F-91128} Palaiseau Cedex, France}
\end{center}\end{sloppypar}
\vspace{2mm}
\begin{sloppypar}
\noindent
V.~Ciulli,
E.~Focardi,
G.~Parrini
\nopagebreak
\begin{center}
\parbox{15.5cm}{\sl\samepage
Dipartimento di Fisica, Universit\`a di Firenze, INFN Sezione di Firenze,
I-50125 Firenze, Italy}
\end{center}\end{sloppypar}
\vspace{2mm}
\begin{sloppypar}
\noindent
A.~Antonelli,
M.~Antonelli,
G.~Bencivenni,
F.~Bossi,
G.~Capon,
F.~Cerutti,
V.~Chiarella,
P.~Laurelli,
G.~Mannocchi,$^{5}$
G.P.~Murtas,
L.~Passalacqua
\nopagebreak
\begin{center}
\parbox{15.5cm}{\sl\samepage
Laboratori Nazionali dell'INFN (LNF-INFN), I-00044 Frascati, Italy}
\end{center}\end{sloppypar}
\vspace{2mm}
\begin{sloppypar}
\noindent
J.~Kennedy,
J.G.~Lynch,
P.~Negus,
V.~O'Shea,
A.S.~Thompson
\nopagebreak
\begin{center}
\parbox{15.5cm}{\sl\samepage
Department of Physics and Astronomy, University of Glasgow, Glasgow G12
8QQ,United Kingdom$^{10}$}
\end{center}\end{sloppypar}
\vspace{2mm}
\begin{sloppypar}
\noindent
S.~Wasserbaech
\nopagebreak
\begin{center}
\parbox{15.5cm}{\sl\samepage
Utah Valley State College, Orem, UT 84058, U.S.A.}
\end{center}\end{sloppypar}
\vspace{2mm}
\begin{sloppypar}
\noindent
R.~Cavanaugh,$^{4}$
S.~Dhamotharan,$^{21}$
C.~Geweniger,
P.~Hanke,
V.~Hepp,
E.E.~Kluge,
A.~Putzer,
H.~Stenzel,
K.~Tittel,
M.~Wunsch$^{19}$
\nopagebreak
\begin{center}
\parbox{15.5cm}{\sl\samepage
Kirchhoff-Institut f\"ur Physik, Universit\"at Heidelberg, D-69120
Heidelberg, Germany$^{16}$}
\end{center}\end{sloppypar}
\vspace{2mm}
\begin{sloppypar}
\noindent
R.~Beuselinck,
W.~Cameron,
G.~Davies,
P.J.~Dornan,
M.~Girone,$^{1}$
N.~Marinelli,
J.~Nowell,
S.A.~Rutherford,
J.K.~Sedgbeer,
J.C.~Thompson,$^{14}$
R.~White
\nopagebreak
\begin{center}
\parbox{15.5cm}{\sl\samepage
Department of Physics, Imperial College, London SW7 2BZ,
United Kingdom$^{10}$}
\end{center}\end{sloppypar}
\vspace{2mm}
\begin{sloppypar}
\noindent
V.M.~Ghete,
P.~Girtler,
E.~Kneringer,
D.~Kuhn,
G.~Rudolph
\nopagebreak
\begin{center}
\parbox{15.5cm}{\sl\samepage
Institut f\"ur Experimentalphysik, Universit\"at Innsbruck, A-6020
Innsbruck, Austria$^{18}$}
\end{center}\end{sloppypar}
\vspace{2mm}
\begin{sloppypar}
\noindent
E.~Bouhova-Thacker,
C.K.~Bowdery,
D.P.~Clarke,
G.~Ellis,
A.J.~Finch,
F.~Foster,
G.~Hughes,
R.W.L.~Jones,
M.R.~Pearson,
N.A.~Robertson,
M.~Smizanska
\nopagebreak
\begin{center}
\parbox{15.5cm}{\sl\samepage
Department of Physics, University of Lancaster, Lancaster LA1 4YB,
United Kingdom$^{10}$}
\end{center}\end{sloppypar}
\vspace{2mm}
\begin{sloppypar}
\noindent
O.~van~der~Aa,
C.~Delaere,$^{28}$
G.Leibenguth,$^{31}$
V.~Lemaitre$^{29}$
\nopagebreak
\begin{center}
\parbox{15.5cm}{\sl\samepage
Institut de Physique Nucl\'eaire, D\'epartement de Physique, Universit\'e Catholique de Louvain, 1348 Louvain-la-Neuve, Belgium}
\end{center}\end{sloppypar}
\vspace{2mm}
\begin{sloppypar}
\noindent
U.~Blumenschein,
F.~H\"olldorfer,
K.~Jakobs,
F.~Kayser,
A.-S.~M\"uller,
B.~Renk,
H.-G.~Sander,
S.~Schmeling,
H.~Wachsmuth,
C.~Zeitnitz,
T.~Ziegler
\nopagebreak
\begin{center}
\parbox{15.5cm}{\sl\samepage
Institut f\"ur Physik, Universit\"at Mainz, D-55099 Mainz, Germany$^{16}$}
\end{center}\end{sloppypar}
\vspace{2mm}
\begin{sloppypar}
\noindent
A.~Bonissent,
P.~Coyle,
C.~Curtil,
A.~Ealet,
D.~Fouchez,
P.~Payre,
A.~Tilquin
\nopagebreak
\begin{center}
\parbox{15.5cm}{\sl\samepage
Centre de Physique des Particules de Marseille, Univ M\'editerran\'ee,
IN$^{2}$P$^{3}$-CNRS, F-13288 Marseille, France}
\end{center}\end{sloppypar}
\vspace{2mm}
\begin{sloppypar}
\noindent
F.~Ragusa
\nopagebreak
\begin{center}
\parbox{15.5cm}{\sl\samepage
Dipartimento di Fisica, Universit\`a di Milano e INFN Sezione di
Milano, I-20133 Milano, Italy.}
\end{center}\end{sloppypar}
\vspace{2mm}
\begin{sloppypar}
\noindent
A.~David,
H.~Dietl,$^{32}$
G.~Ganis,$^{27}$
K.~H\"uttmann,
G.~L\"utjens,
W.~M\"anner$^{32}$,
\mbox{H.-G.~Moser},
R.~Settles,
M.~Villegas,
G.~Wolf
\nopagebreak
\begin{center}
\parbox{15.5cm}{\sl\samepage
Max-Planck-Institut f\"ur Physik, Werner-Heisenberg-Institut,
D-80805 M\"unchen, Germany\footnotemark[16]}
\end{center}\end{sloppypar}
\vspace{2mm}
\begin{sloppypar}
\noindent
J.~Boucrot,
O.~Callot,
M.~Davier,
L.~Duflot,
\mbox{J.-F.~Grivaz},
Ph.~Heusse,
A.~Jacholkowska,$^{6}$
L.~Serin,
\mbox{J.-J.~Veillet}
\nopagebreak
\begin{center}
\parbox{15.5cm}{\sl\samepage
Laboratoire de l'Acc\'el\'erateur Lin\'eaire, Universit\'e de Paris-Sud,
IN$^{2}$P$^{3}$-CNRS, F-91898 Orsay Cedex, France}
\end{center}\end{sloppypar}
\vspace{2mm}
\begin{sloppypar}
\noindent
P.~Azzurri, 
G.~Bagliesi,
T.~Boccali,
L.~Fo\`a,
A.~Giammanco,
A.~Giassi,
F.~Ligabue,
A.~Messineo,
F.~Palla,
G.~Sanguinetti,
A.~Sciab\`a,
G.~Sguazzoni,
P.~Spagnolo,
R.~Tenchini,
A.~Venturi,
P.G.~Verdini
\samepage
\begin{center}
\parbox{15.5cm}{\sl\samepage
Dipartimento di Fisica dell'Universit\`a, INFN Sezione di Pisa,
e Scuola Normale Superiore, I-56010 Pisa, Italy}
\end{center}\end{sloppypar}
\vspace{2mm}
\begin{sloppypar}
\noindent
O.~Awunor,
G.A.~Blair,
G.~Cowan,
A.~Garcia-Bellido,
M.G.~Green,
T.~Medcalf,$^{25}$
A.~Misiejuk,
J.A.~Strong,
P.~Teixeira-Dias
\nopagebreak
\begin{center}
\parbox{15.5cm}{\sl\samepage
Department of Physics, Royal Holloway \& Bedford New College,
University of London, Egham, Surrey TW20 OEX, United Kingdom$^{10}$}
\end{center}\end{sloppypar}
\vspace{2mm}
\begin{sloppypar}
\noindent
R.W.~Clifft,
T.R.~Edgecock,
P.R.~Norton,
I.R.~Tomalin,
J.J.~Ward
\nopagebreak
\begin{center}
\parbox{15.5cm}{\sl\samepage
Particle Physics Dept., Rutherford Appleton Laboratory,
Chilton, Didcot, Oxon OX11 OQX, United Kingdom$^{10}$}
\end{center}\end{sloppypar}
\vspace{2mm}
\begin{sloppypar}
\noindent
\mbox{B.~Bloch-Devaux},
D.~Boumediene,
P.~Colas,
B.~Fabbro,
E.~Lan\c{c}on,
\mbox{M.-C.~Lemaire},
E.~Locci,
P.~Perez,
J.~Rander,
B.~Tuchming,
B.~Vallage
\nopagebreak
\begin{center}
\parbox{15.5cm}{\sl\samepage
CEA, DAPNIA/Service de Physique des Particules,
CE-Saclay, F-91191 Gif-sur-Yvette Cedex, France$^{17}$}
\end{center}\end{sloppypar}
\vspace{2mm}
\begin{sloppypar}
\noindent
A.M.~Litke,
G.~Taylor
\nopagebreak
\begin{center}
\parbox{15.5cm}{\sl\samepage
Institute for Particle Physics, University of California at
Santa Cruz, Santa Cruz, CA 95064, USA$^{22}$}
\end{center}\end{sloppypar}
\vspace{2mm}
\begin{sloppypar}
\noindent
C.N.~Booth,
S.~Cartwright,
F.~Combley,$^{25}$
P.N.~Hodgson,
M.~Lehto,
L.F.~Thompson
\nopagebreak
\begin{center}
\parbox{15.5cm}{\sl\samepage
Department of Physics, University of Sheffield, Sheffield S3 7RH,
United Kingdom$^{10}$}
\end{center}\end{sloppypar}
\vspace{2mm}
\begin{sloppypar}
\noindent
A.~B\"ohrer,
S.~Brandt,
C.~Grupen,
J.~Hess,
A.~Ngac,
G.~Prange
\nopagebreak
\begin{center}
\parbox{15.5cm}{\sl\samepage
Fachbereich Physik, Universit\"at Siegen, D-57068 Siegen, Germany$^{16}$}
\end{center}\end{sloppypar}
\vspace{2mm}
\begin{sloppypar}
\noindent
C.~Borean,
G.~Giannini
\nopagebreak
\begin{center}
\parbox{15.5cm}{\sl\samepage
Dipartimento di Fisica, Universit\`a di Trieste e INFN Sezione di Trieste,
I-34127 Trieste, Italy}
\end{center}\end{sloppypar}
\vspace{2mm}
\begin{sloppypar}
\noindent
H.~He,
J.~Putz,
J.~Rothberg
\nopagebreak
\begin{center}
\parbox{15.5cm}{\sl\samepage
Experimental Elementary Particle Physics, University of Washington, Seattle,
WA 98195 U.S.A.}
\end{center}\end{sloppypar}
\vspace{2mm}
\begin{sloppypar}
\noindent
S.R.~Armstrong,
K.~Berkelman,
K.~Cranmer,
D.P.S.~Ferguson,
Y.~Gao,$^{13}$
S.~Gonz\'{a}lez,
O.J.~Hayes,
H.~Hu,
S.~Jin,
J.~Kile,
P.A.~McNamara III,
J.~Nielsen,
Y.B.~Pan,
\mbox{J.H.~von~Wimmersperg-Toeller}, 
W.~Wiedenmann,
J.~Wu,
Sau~Lan~Wu,
X.~Wu,
G.~Zobernig
\nopagebreak
\begin{center}
\parbox{15.5cm}{\sl\samepage
Department of Physics, University of Wisconsin, Madison, WI 53706,
USA$^{11}$}
\end{center}\end{sloppypar}
\vspace{2mm}
\begin{sloppypar}
\noindent
G.~Dissertori
\nopagebreak
\begin{center}
\parbox{15.5cm}{\sl\samepage
Institute for Particle Physics, ETH H\"onggerberg, 8093 Z\"urich,
Switzerland.}
\end{center}\end{sloppypar}
}
\footnotetext[1]{Also at CERN, 1211 Geneva 23, Switzerland.}
\footnotetext[2]{Now at Fermilab, PO Box 500, MS 352, Batavia, IL 60510, USA}
\footnotetext[3]{Also at Dipartimento di Fisica di Catania and INFN Sezione di
 Catania, 95129 Catania, Italy.}
\footnotetext[4]{Now at University of Florida, Department of Physics, Gainesville, Florida 32611-8440, USA}
\footnotetext[5]{Also IFSI sezione di Torino, INAF, Italy.}
\footnotetext[6]{Also at Groupe d'Astroparticules de Montpellier, Universit\'{e} de Montpellier II, 34095, Montpellier, France.}
\footnotetext[7]{Supported by CICYT, Spain.}
\footnotetext[8]{Supported by the National Science Foundation of China.}
\footnotetext[9]{Supported by the Danish Natural Science Research Council.}
\footnotetext[10]{Supported by the UK Particle Physics and Astronomy Research
Council.}
\footnotetext[11]{Supported by the US Department of Energy, grant
DE-FG0295-ER40896.}
\footnotetext[12]{Now at Departement de Physique Corpusculaire, Universit\'e de
Gen\`eve, 1211 Gen\`eve 4, Switzerland.}
\footnotetext[13]{Also at Department of Physics, Tsinghua University, Beijing, The People's Republic of China.}
\footnotetext[14]{Supported by the Leverhulme Trust.}
\footnotetext[15]{Permanent address: Universitat de Barcelona, 08208 Barcelona,
Spain.}
\footnotetext[16]{Supported by Bundesministerium f\"ur Bildung
und Forschung, Germany.}
\footnotetext[17]{Supported by the Direction des Sciences de la
Mati\`ere, C.E.A.}
\footnotetext[18]{Supported by the Austrian Ministry for Science and Transport.}
\footnotetext[19]{Now at SAP AG, 69185 Walldorf, Germany}
\footnotetext[20]{Now at Groupe d' Astroparticules de Montpellier, Universit\'e de Montpellier II, 34095 Montpellier, France.}
\footnotetext[21]{Now at BNP Paribas, 60325 Frankfurt am Mainz, Germany}
\footnotetext[22]{Supported by the US Department of Energy,
grant DE-FG03-92ER40689.}
\footnotetext[23]{Now at Institut Inter-universitaire des hautes Energies (IIHE), CP 230, Universit\'{e} Libre de Bruxelles, 1050 Bruxelles, Belgique}
\footnotetext[24]{Now at Dipartimento di Fisica e Tecnologie Relative, Universit\`a di Palermo, Palermo, Italy.}
\footnotetext[25]{Deceased.}
\footnotetext[26]{Now at SLAC, Stanford, CA 94309, U.S.A}
\footnotetext[27]{Now at CERN, 1211 Geneva 23, Switzerland}
\footnotetext[28]{Research Fellow of the Belgium FNRS}
\footnotetext[29]{Research Associate of the Belgium FNRS} 
\footnotetext[30]{Now at Liverpool University, Liverpool L69 7ZE, United Kingdom} 
\footnotetext[31]{Supported by the Federal Office for Scientific, Technical and Cultural Affairs through
the Interuniversity Attraction Pole P5/27} 
\footnotetext[32]{Now at Henryk Niewodnicznski Institute of Nuclear Physics, Polish Academy of Sciences, Cracow, Poland}   
\setlength{\parskip}{\saveparskip}
\setlength{\textheight}{\savetextheight}
\setlength{\topmargin}{\savetopmargin}
\setlength{\textwidth}{\savetextwidth}
\setlength{\oddsidemargin}{\saveoddsidemargin}
\setlength{\topsep}{\savetopsep}
\normalsize
\newpage
\pagestyle{plain}
\setcounter{page}{1}

\section{Introduction}
\label{sec:intro}

A Higgs model~\cite{higgs} incorporating two doublets of complex
scalar fields~\cite{TwoHiggsDoublets} generates five scalar Higgs bosons, 
three of which are neutral. In some types of models, for certain choices of parameters, 
one of these neutral scalars provides mass only to the fermions and another
couples exclusively to the bosons, i.e. is a ``fermiophobic'' Higgs boson.
Anomalous couplings in the Higgs sector can also enhance the bosonic 
branching fraction~\cite{AnomalousCouplings}.

\begin{figure}
\begin{center}
  \includegraphics[width=7.5cm]{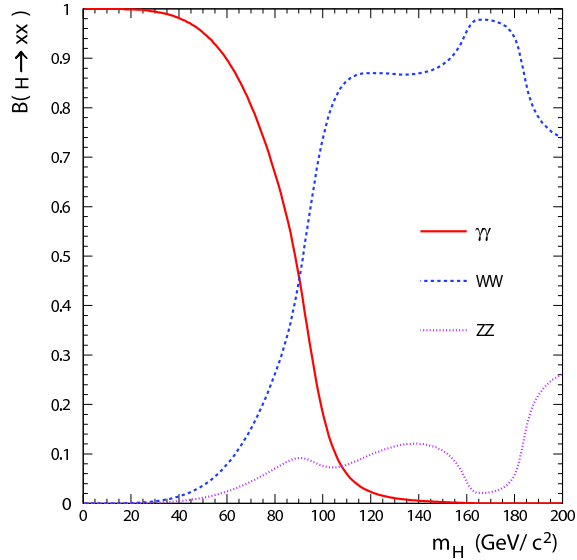}
\end{center}
  \caption{\small Branching fraction of benchmark fermiophobic Higgs boson 
      (defined in the text) into boson pairs as calculated by {\tt HZHA}~\cite{HZHA}.}
  \label{fig:branching}
 \end{figure}

The search for a fermiophobic Higgs boson has been primarily carried out 
by the four LEP experiments in the $\mathrm{H} \to \gamma\gamma$ channel, 
in which the Higgs boson couples to photons via a W loop~\cite{Aleph_hgg,Delphi_hgg,L3_hgg,Opal_hgg}. 
A benchmark fermiophobic Higgs boson is defined by considering Standard-Model-like couplings to bosons, and null couplings to fermions.
Current analyses exclude the benchmark fermiophobic Higgs boson 
up to a mass of~109.7\gevct~\cite{LEPggpaper}.
For fermiophobic Higgs bosons heavier than 90\gevct, the predicted 
$\mathrm{H} \to \gamma\gamma$ branching ratio becomes small relative to 
the predicted $\mathrm{H \to WW}$  branching ratio (Fig.~\ref{fig:branching}) 
motivating a search in this new channel. 
Such an analysis has already been carried out by the L3 collaboration~\cite{L3_hWW}
and is performed here with data collected by the ALEPH detector.

The main production processes at e$^+$e$^-$ colliders for a fermiophobic Higgs 
boson are $\mathrm{e^+e^- \to Z^* \to ZH}$ (Higgsstrahlung), WW and ZZ fusion.
The cross sections of the boson fusion production processes are considerably 
smaller than that of the Higgsstrahlung process at LEP centre-of-mass (CM) energies.
In the mass range kinematically accessible for Higgsstrahlung at LEP, 
one of the virtual W bosons is expected to be near on-shell, and 
the other (denoted $\mathrm W^*$) to have a much smaller mass and energy.
In this paper, all the signatures originating from the $\mathrm{Z\to q\bar{q}, \nu\bar{\nu}, 
\ell^+\ell^-}$ decays and the $\mathrm{W,W^*\to q\bar{q}',\ell^\pm\nu}$ decays 
are searched for. For simplicity and conciseness, the term ``lepton'' (and the 
corresponding symbol $\ell$) refers to electrons and muons only. Leptonic tau 
decays are not specifically addressed, but the corresponding selected events are 
included in the final results. The hadronic tau decays in 
$\mathrm{e^+e^- \to ZWW^* \to \ell^+\ell^-\tau\nu q\bar{q}'}$ are also looked for.
The analysis is performed on the data taken in the years 1999 and 2000 at 
CM energies ranging from 191 to 209 \gev. 
The luminosities and CM energies are shown in Table~\ref{tab:lumi}.

\begin{table}[b]
\begin{center}
\caption{\small Integrated luminosities and centre-of-mass (CM) energy
   ranges for the data collected by the ALEPH detector for the years 1999 and 2000.}
\vskip 2ex
\begin{tabular}{|c|c|c|} 
 \hline 
\rule{0pt}{4.6mm}
Year & Luminosity  (\pbm) & CM energy (\gev) \\ \hline
2000 &   $7.3\pm0.05$     & $207-209$        \\
      & $125.9\pm0.6$     & $206-207$        \\
      &  $81.4\pm0.4$     & $204-206$        \\ 
\hline
 1999 &  $42.6\pm0.2$     & $201-203$        \\
      &  $87.2\pm0.4$     & 199.5            \\
      &  $79.9\pm0.4$     & 195.5            \\
      &  $28.9\pm0.1$     & 191.6            \\
\hline
\end{tabular}
\label{tab:lumi}
\end{center}
\end{table}

This paper is organized as follows. A brief description of the ALEPH detector 
is given is Section~\ref{sec:detector}. The signal and 
background simulations are summarized in Section~\ref{sec:monte}. 
The overall search strategy is presented in Section~\ref{sec:strategy} and 
the specific selection algorithms in Section~\ref{sec:selec}. 
The results are reported in Section~\ref{sec:res}.

\section{The ALEPH detector}
\label{sec:detector}

A detailed description of the ALEPH detector can be found in 
Ref.~\cite{ALEPH}, and of its performance in 
Ref.~\cite{Performance}. Here, only a brief description of the detector
elements and the algorithms relevant to this analysis is given. 

The trajectories of charged particles are measured with a silicon vertex 
detector (VDET), a cylindrical drift chamber (ITC) and a large time 
projection chamber (TPC),  all immersed in a 1.5\,T axial magnetic field 
provided by a superconducting  solenoidal coil. The energy of electrons, 
photons, and hadrons is measured with the electromagnetic (ECAL), the 
hadron (HCAL) and the luminosity (LCAL and SiCAL) calorimeters. The ECAL, 
placed between the TPC and the coil, is a highly segmented calorimeter, 
which is used to identify electrons and photons, and to measure their
energy and position. The LCAL and SiCAL extend the calorimetric coverage
down to 34\,mrad from the beam axis. The HCAL consists of an instrumented
iron return yoke. It provides the measurement of hadronic energy and, 
together with external chambers, muon identification.

In the following, {\it good} tracks (or simply tracks) are defined as 
charged particle tracks reconstructed with at least four hits in the time 
projection chamber, originating from within a cylinder of 20\,cm length
and 2\,cm radius, coaxial with the beam and centred on the nominal
collision point, and with a polar angle with respect to the beam
such that $\vert \cos\theta \vert < 0.95$. 

In this analysis, all searches make use of the same lepton identification 
criteria, where needed and applicable. Electrons are identified by comparing
the momentum measured in the tracking detectors with the energy measured in the
ECAL, by the depth and shape of the ECAL shower, and by the specific 
ionization information from the TPC, when available. Muons are identified by 
their characteristic hit pattern in the hadron calorimeter, and must have 
at least one associated hit in the muon chambers. Lepton identification is 
described in detail in Ref.~\cite{HFLepId}.

Global event quantities, such as the total visible energy (\esum), 
the total visible mass (\msum), or the missing energy 
(\emiss) are measured with an energy-flow algorithm~\cite{Performance}, which 
combines individual tracker and calorimeter measurements into energy-flow 
``particles''. These objects are classified as photons, electrons, muons, 
neutral or charged hadrons. The jets used in the present analysis are obtained by clustering the energy-flow particles with the Durham jet-finding
algorithm~\cite{durham}, and allow various event topological variables to be 
determined. These include, for example, the acollinearity angle \acolin (respectively the acoplanarity angle \acopl) between two 
jets (respectively between their projections onto the plane transverse to the 
beam axis) in an event forced to form two jets, the transition values 
$y_{ij}$ of the resolution parameter $y_{\rm cut}$ at which the number of 
jets in an event switches from $j$ to $i$ jets, or the good track 
multiplicity $N_{{\rm ch}, i}$ in jet $i$. Other specific variables, 
related either to the event topology, to the jet properties, or to 
the lepton characteristics, are described in turn in Section~\ref{sec:selec}.

\section{Simulated samples}
\label{sec:monte}

Signal samples were generated using {\tt HZHA}~\cite{HZHA}
for Higgs boson masses between 90 and 117\gevct\ and at seven different 
CM energies: 191.6, 195.5, 199.5, 201.6, 204.9, 206.5 and 208\gev,
including Higgsstrahlung and fusion processes.
All decays of the Z and W bosons were considered.
In the {\tt HZHA} code, there is no spin correlation between the W bosons 
coming from the Higgs boson decay. The signal events were therefore 
re-weighted to take into account this spin correlation. 
The event weight was computed as the ratio between the full 
four-fermion matrix element and the {\tt HZHA} matrix element~\cite{tomek}.

Event samples of all Standard Model (SM) background processes relevant for 
the Higgs boson search were also generated: 
the Bhabha process was simulated with {\tt BHWIDE~1.01} \cite{bhwide}, 
$\mathrm{q\bar q}$, dimuon and ditau events with {\tt KK~4.14}~\cite{kk2f},
\gaga\ processes with {\tt PHOT02}~\cite{phot02}.
In the following, these processes will be grouped under the label ``two-fermions''. 
WW production was simulated with {\tt KORALW~1.51}~\cite{koralw}
and the remaining four-fermion processes with {\tt PYTHIA 6.1}~\cite{pythia}.
The background event samples were generated at the same CM energy values 
as the signal.
The simulated sample sizes are at least a factor 20 greater than the data.
A detailed simulation of the detector response was applied to both background and signal events.


\section{Search strategy}
\label{sec:strategy}

\subsection{Event classes, topological searches and targeted channels}
\label{sec:class}
The event selection is subdivided in a number of topological searches, each 
of which targets a specific final state (or channel) arising from 
the ${\rm Z W W^\ast}$  production. The list of channels addressed in this
paper is given in the second column of Table~\ref{tab:class}, with the Z, the W
and the ${\rm W^\ast}$ decays given in this order, together with the 
corresponding branching fractions. Altogether, almost 80\% of the possible 
final states are targeted by the selection algorithms developed for this 
study.  

Events are first separated in four exclusive classes according to the number 
and the energy of identified leptons in the final state. The four classes, 
further subdivided in different topologies (or subclasses), are 
defined as follows, and are displayed in the first column of 
Table~\ref{tab:class}.

\begin{itemize}

\item Fully Hadronic  (Class 1): in this class, only events with neither 
energetic nor isolated identified leptons are selected. It addresses 
final states exclusively with 
hadronic jets, with (class 1a) or without (class 1b) 
missing energy, depending on whether the Z decays in a pair of quarks or 
a pair of neutrinos.  

\item Two Hard Leptons (Class 2): in this class, events with at least 
two energetic identified leptons are selected. It addresses final states 
in which the Z decays into a lepton pair. Five different topological searches  
were developed according to the W and ${\rm W^\ast}$ decay modes. 

\item One Hard Lepton (class 3): in this class, events with exactly one 
energetic identified lepton are selected. It addresses final states in which 
the W decays to $\ell\nu$. Four different topologies were defined according 
to the hadronic activity and the missing energy, to target the remaining Z 
and ${\rm W^\ast}$ decays.

\item One Soft Lepton (class 4): in this class, events with exactly one 
isolated identified lepton that does not meet the momentum requirement 
of class 3 are selected, to address the ${\rm W}^\ast\to \ell\nu$ decays. 
Two different topological searches were developed according to the 
hadronic activity and the missing energy, to address the hadronic 
and invisible Z decays.
\end{itemize} 

These definitions, and the corresponding topological searches, were developed to minimize the cross-channel contamination between the different subclasses.

\begin{table}[b!]
\begin{center}
\caption{\small Search topologies and targeted final states (with 
corresponding branching fractions) in the four event classes. For the 
targeted final states, the Z, W, and ${\rm W^\ast}$ decays are given 
in this order.}
\vskip 2ex
\begin{tabular}{| l | l r |} \hline
\rule{0pt}{4.6mm}
 {\bf Class} and topology                     & Targeted Channel                          &    (BR)   \\ 
\hline\hline
 {\bf 1: Fully-Hadronic}                 & {\bf No leptonic decay}                   & {\bf (0.422)}  \\
\hline
 1a: 6 jets                              & ${\rm q\bar q \, q\bar q \, q\bar q}$     & (0.328)  \\
 1b: 4 jets and \emiss\                  & $\nu\bar\nu\,{\rm q\bar q\, q\bar q}$     & (0.094)  \\ 
\hline\hline
 {\bf 2: Two-Hard-Leptons}               & {\bf Z leptonic decays}                   & {\bf (0.054)}  \\           
\hline
2a: plus jets                            & $\ell^+\ell^-\,{\rm q\bar q\,q\bar q}$    & (0.032)  \\
2t: plus jets and \emiss\                & $\ell^+\ell^-\,\tau\nu\,{\rm q\bar q}$    & (0.003)  \\
2b: plus jets and 1 soft lepton          & $\ell^+\ell^-\,{\rm q\bar q}\,\ell\nu$    & (0.010)  \\
2c: plus jets and 1 hard lepton          & $\ell^+\ell^-\,\ell\nu \,{\rm q\bar q}$   & (0.007)  \\
2d: plus 1 hard lepton and 1 track       & $\ell^+\ell^-\,\ell\nu  \,\ell\nu $       & (0.003)  \\ 
\hline\hline
{\bf 3: One-Hard-Lepton} (and $E_{\rm miss}$) & {\bf W leptonic decays }               & {\bf (0.171)} \\
\hline
3a: plus jets                            & ${\rm q\bar q}\,\ell\nu \,{\rm q\bar q}$  & (0.101)  \\
3b: plus jets and 1 soft lepton          & ${\rm q\bar q}\,\ell\nu \,\ell\nu $       & (0.031)  \\
3c: plus 1 track and \mmiss\             & $\nu\bar\nu\,\ell\nu  \,\ell\nu$          & (0.029)  \\
3d: plus jets and \mmiss\                & $\nu\bar\nu\,\ell\nu\,\rm{q\bar q}$       & (0.008)  \\
\hline\hline
 {\bf 4: One-Soft-Lepton}                  & {\bf ${\rm W^\ast}$ leptonic decays}     & {\bf (0.130)}  \\
\hline
4a: plus jets                            & ${\rm q\bar q\,q\bar q}\,\ell\nu$        & (0.101)  \\
4b: plus jets and \mmiss\                & $\nu\bar\nu\,{\rm q\bar q}\,\ell\nu$     & (0.029)  \\ 
\hline
\end{tabular}
\label{tab:class}
\end{center}
\end{table}

\subsection{Topological search optimization}
\label{sec:optimization}
In each of the topological searches, the selection criteria were tailored
to optimize the combined sensitivity to a Higgs boson mass hypothesis of 
$110\gevct$, which is near the expected experimental sensitivity in the 
fermiophobic scenario. To do so, the expected combined confidence level 
on the signal hypothesis, \ecls, that would be obtained on average if no
signal were present, is minimized~\cite{optim} with respect to the position of 
the cuts on most of the selection variables. An estimate of the value of 
\ecls\ is determined with a toy Monte Carlo method using the approximate 
formula of Ref.~\cite{jinMcnamara}, with the algorithm of Ref.~\cite{junk}. 

For each topology, the determination of \ecls\ requires the expected number of background events ($N_{\rm b}$), the number of signal events expected from 
the targeted final state $N_{\rm s}$, and the expected distribution of 
a variable $D$ aimed at discriminating between the signal and the background.
The observed values of the confidence levels on the signal and on the 
background hypotheses, \cls\ and \clb, are obtained in the same 
way from the number of events observed ($N_{\rm d}$) and the value of 
$D$ for each of these events.

In the optimization process the values of $N_{\rm b}$ and $N_{\rm s}$ 
correspond to the CM energies and the integrated luminosity 
collected in the year 2000. Lower energy data sets would indeed not 
contribute significantly to the combined sensitivity to a $110\gevct$
Higgs boson signal, and are therefore absent from all distributions 
presented in Section~\ref{sec:selec}. Since, however, they increase 
the sensitivity to smaller masses, the data taken in 1999 were included in 
the final result (Section~\ref{sec:res}).

\section{Event selection}
\label{sec:selec}

\subsection{Preselection and class assignment}
\label{sec:presel}

Common preselection cuts are applied in all four classes in order to strongly reduce the $\gamma \gamma$ and $\ell^+\ell^-$ backgrounds. 
The energy within 12\textdegree\ of the beam axis, \ef, must be less than 40\% of the CM energy. 
The acollinearity must be less than 170\textdegree\ for events with less than four tracks. 
Finally, the total invariant mass \mtot\ and total transverse momentum \pt\ of the event must satisfy 
$\msum + 6 \ptsum > 0.2 \sqrts$. 

The event-to-(sub)class assignment is based on the energies $E_{\ell_i}$ and the  isolations $I_{\ell_i} = 1 - \cos \theta_{\ell_i T}$ ($i=1,2,3$) of the three most energetic leptons in the event. Here $\theta_{\ell_i T}$ is the angle between the $i$th lepton direction and the closest track in the event. If less than three leptons are found, the corresponding energies and isolation variables are set to 0 and $10^{-20}$ respectively.
Events in which the most energetic lepton is ``hard'' ($\ELo > 25\gev$) are assigned to class 2 or 3 depending on the energy of the second most energetic lepton.
To separate the remaining events between classes 1 and 4, a linear discriminant $D_{14}$ is built with \ELo, the total missing three-momentum (\pmiss), and the isolation of the most energetic lepton (\ILo):
\begin{equation}
  D_{14} = 2.3 \ELo + \pmiss + 4.8 \ln(\ILo).
\end{equation}
Details of the partition process are presented in Table~\ref{tab:subclasses}.
The criteria used to define all the subclasses presented in Section~\ref{sec:class} are also shown in the table.
Subclasses 1a and 1b are separated by a cut on the missing mass $\mmiss$. 
Subclasses 2a and 2t are separated from subclasses 2b, 2c and 2d with a cut on \ELth. 
Subclass 2b is finally identified with a cut on the total hadronic energy, $\mathrm{E_{had}}$.
Subclasses 2c and 2d are separated from each other by a cut on the number of tracks 
(\nch) while subclasses 2a and 2t are distinguished using the hadronic activity.
The separation between subclasses 3a, 3b and 3c, 3d is achieved by cutting on the total and missing invariant mass of the event.
A cut on the energy of the second most energetic lepton is used to distinguish subclasses 3a, 3d from 3b, 3c.
Finally, subclasses 4a and 4b are separated from each other by cutting on the total invariant mass of the event.

\begin{table}[b!]
\begin{center}
\caption{\small Details of the partition of events into each of the thirteen subclasses. Energies are expressed in \gev\ and masses in \gevct. }
\vskip 2ex
\begin{tabular}{| c | c c c c c c c c |} \hline
\rule{0pt}{4.6mm}
\raisebox{-2ex}{Subclass} & \multicolumn{8}{|c|}{\raisebox{-0.5ex}{Selection Criteria}} \\
\cline{2-9} 
       & \ELo  & \ELt      & \ELth & $D_{14}$ & \msumf & \mmiss & \ehad & \nch \\
\hline
1a     & $<25$ &           &       &  $<13$   &        & $<60$  &       &      \\
1b     & $<25$ &           &       &  $<13$   &        & $>60$  &       &      \\
\hline
2a     & $>25$ & $>20$     & $<8$  &          &        &        & $>60$ &      \\
2t     & $>25$ & $>20$     & $<8$  &          &        &        & $<60$ &      \\
2b     & $>25$ & $>20$     & $>8$  &          &        &        & $>60$ &      \\
2c     & $>25$ & $>20$     & $>8$  &          &        &        & $<60$ & $>4$ \\
2d     & $>25$ & $>20$     & $>8$  &          &        &        & $<60$ & $=4$ \\
\hline
3a     & $>25$ & $<10$     &       &          & $>0.4$ &  $<95$ &       &      \\
3b     & $>25$ & $[10,20]$ &       &          & $>0.4$ &  $<95$ &       &      \\
3c     & $>25$ & $[10,20]$ &       &          & $<0.4$ &  $>95$ &       &      \\
3d     & $>25$ & $<10$     &       &          & $<0.4$ &  $>95$ &       &      \\
\hline
4a     & $<25$ &           &       &  $>13$   & $>0.6$ &        &       &      \\
4b     & $<25$ &           &       &  $>13$   & $<0.6$ &        &       &      \\
\hline
\end{tabular}
\label{tab:subclasses}
\end{center}
\end{table}

\subsection{Class 1: Fully hadronic final state}
\label{sec:class1}

\subsubsection{Class 1a: Six-jets final state\\{\it targeted channel: } $\mathrm{ ZH\rightarrow q\bar{q}\,q\bar{q}\,q\bar{q}}$}
\label{sec:class1a}

This subclass is characterized by a final state with a large number of tracks, a value of the total mass divided by the CM energy ($\msumf$) close to 1, no missing longitudinal momentum ($\plsumf$), an intermediate  sphericity ($\spher$)~\cite{thrust}, and a high value of $\ln(\ythf)$. 
Corresponding preselection cuts are applied, in addition to those applied in the four classes,
and are presented in Table~\ref{tab:cut_class1}, together with the numbers of selected data, background and signal events.
After the preselection, the background is dominated by WW events. 

The final selection cuts are given in the lower section of Table~\ref{tab:cut_class1}. 
The variables $\fsntmi$, $\fsanmi$ and $\fsanma$ are computed after having forced the event to be clustered into six jets. 
The smallest number of tracks in any jet is denoted $\fsntmi$, and the variables $\fsanmi$ and $\fsanma$ represent the smallest and largest angle between any pair of jets. 
The jet pair with the invariant mass closest to 91.2\gevct\ is assigned to the Z. 
The two least energetic jets are assigned to the $\mathrm W^*$, and the remaining two jets to the 
W. The invariant mass of the Z, W and $\mathrm W^*$ are denoted $M_1$, 
$M_2$ and $M_3$, respectively.

The discriminant variable $\fsmt$ enters the \cls\ computation and is shown in Figs.~\ref{fig:class1_discr}a and~\ref{fig:class1_discr}b, after the preselection and the final selection, respectively.
In this subclass, the targeted-signal efficiency is 61\% and the number of expected background events is 163.7.

\subsubsection{Class 1b: Four-jets and missing transverse momentum final state\\{\it targeted channel: } $\mathrm{ ZH\rightarrow \nu\bar{\nu}\,q\bar{q}\,q\bar{q}}$}
\label{sec:class1b}

This subclass is characterized by a final state with a large missing mass. 
The preselection cuts used for subclass 1a are also relevant with, in general, weaker cut values, 
as shown in Table~\ref{tab:cut_class1}. 
The class 1b preselection has two additional cuts on the value of the cosine of the polar angle 
of the missing momentum \cthmiss\ and on \ef. 
After the preselection, the proportion of main background events is 43\% WW, 30\% q$\bar{\rm q}$, 12\% We$\nu$ and 12\% ZZ. 

The final selection cuts are detailed in the lower section of Table~\ref{tab:cut_class1}.
The acoplanarity and $\ptsumf$ are used as well as the thrust $\thrust$~\cite{thrust} of the event. 
The event is forced to form four jets. The smallest number of tracks in any jet is denoted $\ffntmi$, and the variables $\ffanmi$ and $\ffanma$ represent the smallest and largest angle between any two jets.

\begin{figure}
\begin{center}
\begin{tabular}{cc}
\includegraphics[width=0.44\textwidth]{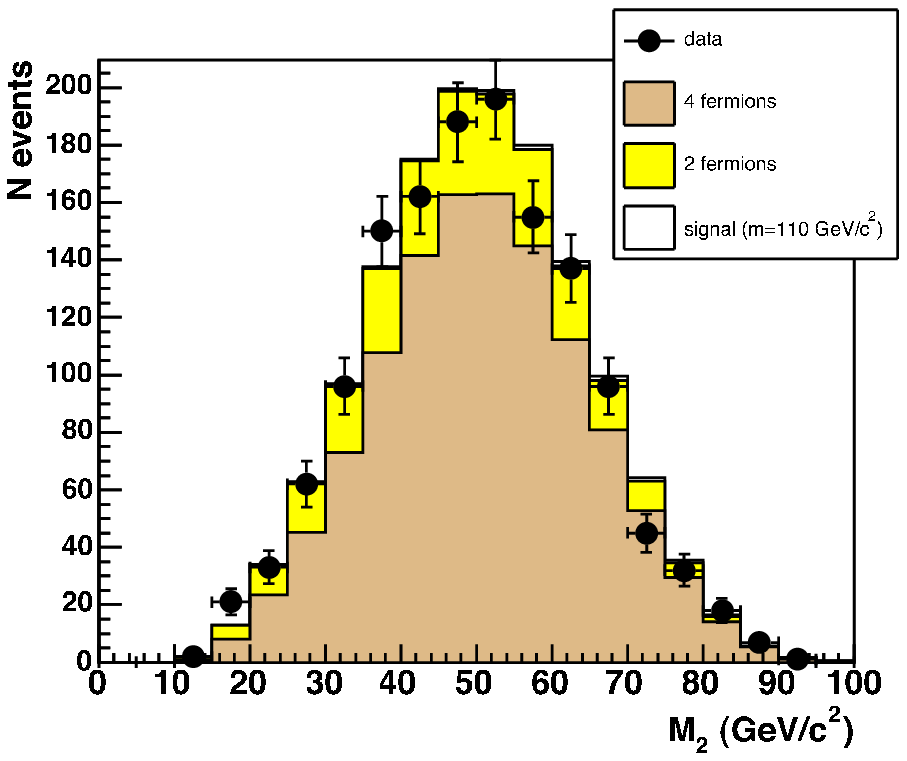} & \includegraphics[width=0.435\textwidth]{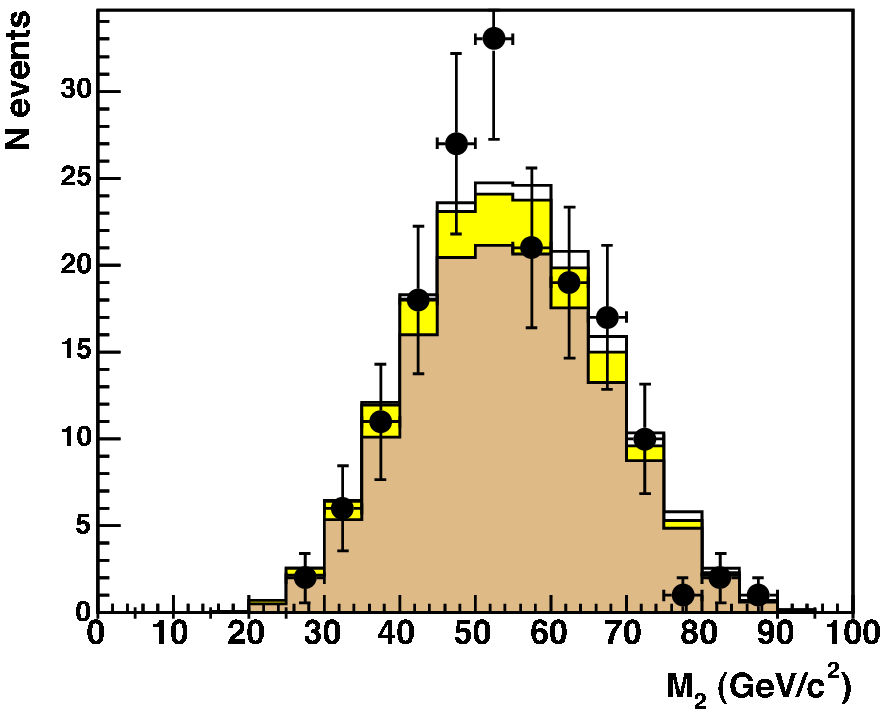} \\
\small (a) & \small (b)\\
\includegraphics[width=0.44\textwidth]{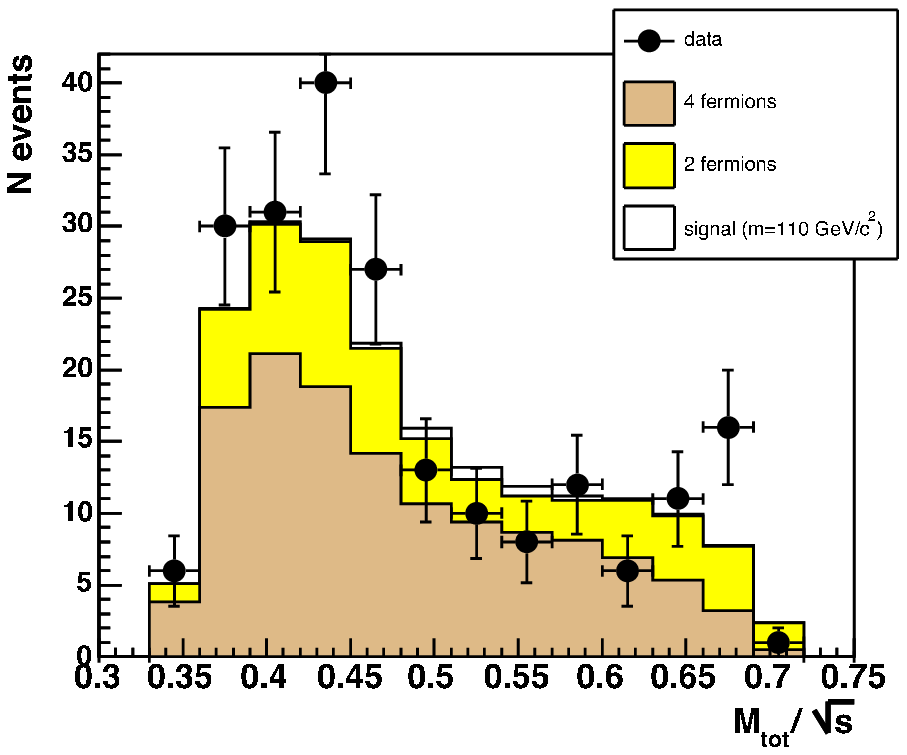} &  \includegraphics[width=0.435\textwidth]{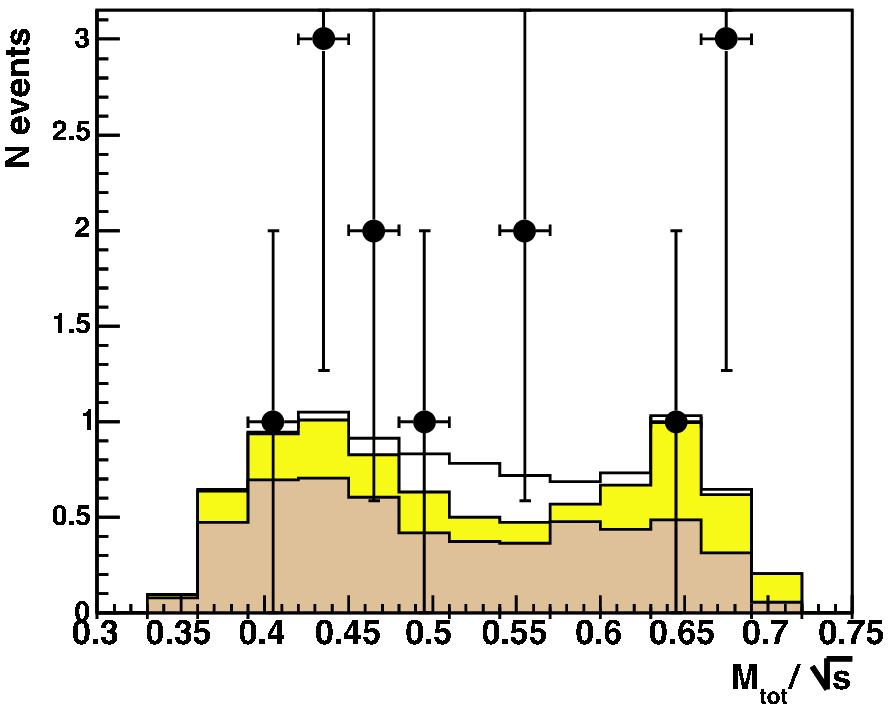}\\
\small (c) & \small (d)\\
\end{tabular}
\put(-85,185){\Large \bf ALEPH}
\caption{\small Discriminant variable $M_2$ for class 1a events after the
        preselection (a) and after the final selection cuts (b). 
        Discriminant variable \msumf\ for class 1b events after the 
        preselection (c) and after the final selection cuts (d). 
	All distributions are obtained from year 2000 data.
        }
\label{fig:class1_discr}
\end{center}
\end{figure}

The discriminant variable $\msumf$ is shown in 
Figs.~\ref{fig:class1_discr}c and~\ref{fig:class1_discr}d after the preselection 
and final selection respectively.
In this subclass, the targeted-signal efficiency is 32\% and the number of background events is 8.2.

\begin{table}[t]
\begin{center}
\caption{\small Selection criteria for each subclass in class 1. The numbers of signal 
             ($N_{\mathrm s}$), background ($N_{\mathrm b}$) and data ($N_{\mathrm d}$) events
             are given in the table for the year 2000. Energies, momenta 
             and masses are expressed in\gev,\gevc~ and\gevct, respectively.}
\vspace*{0.2cm}
\resizebox{\textwidth}{!}{
\hspace{-.3cm}
\begin{tabular}{ll}
\begin{tabular}{|l|p{5cm}|c|c|c|} 
\hline
Subclass 1a  &  Cuts                           & $N_{\mathrm s}$ & $N_{\mathrm b}$ & $N_{\mathrm d}$ \\\hline\hline
Preselection &  $\nch>25$                      &         &         &         \\
             &  $\msumf >0.6$                  &         &         &         \\
             &  $\plsumf<0.15$                 &         &         &         \\
             &  $0.1<\spher<0.85$              &         &         &         \\
             &  $\ln(\ythf)>-5.3$              & 10.9    & 1452.0  & 1401    \\ 
             &                                 &         &         &         \\
             &                                 &         &         &         \\\hline\hline
Topological  &  $\spher>0.13$                  & 10.7    & 1374.1  & 1325    \\
             &  $\nch>32$                      & 9.17    & 866.9   & 845     \\
             &  $\ln(\yfis)>-7.2$              & 8.18    & 494.5   & 481     \\
             &  $\ln(\yot+\ythf+\yfis)>-0.83$  & 8.07    & 476.4   & 456     \\
             &  $\fsntmi>0$                    & 6.28    & 231.5   & 247     \\ 
             &  $(\fsanmi+\fsanma)>190.^\circ$ & 6.21    & 212.3   & 229     \\ \hline
Masses       &  $\fsmo<117.$                    & 6.10    & 186.7   & 200     \\
             &  $\fsmth>13.$                    & 5.74    & 163.7   & 168     \\ 
&&&&\\\hline
\end{tabular}
 &
\begin{tabular}{|l|p{5cm}|l|c|c|c|c|} 
\hline
Subclass 1b  &  Cuts                    & $N_{\mathrm s}$ & $N_{\mathrm b}$ & $N_{\mathrm d}$ \\ \hline \hline
Preselection &  $\cthmiss<0.9$          &         &         &         \\ 
             &  $\ef<0.05$              &         &         &         \\ 
             &  $\nch>12$               &         &         &         \\
             &  $\msumf >0.35$          &         &         &         \\
             &  $\plsumf<0.2$           &         &         &         \\ 
             &  $0.03<\spher<0.8$       &         &         &         \\
             &  $\ln(\ythf) >-7.6$      & 3.60    & 190.4   & 211     \\\hline \hline 
Topological  &  $\nch>24$               & 1.77    & 31.2    & 46      \\
             &  $\ln(\ythf) >-6.6$      & 1.73    & 28.2    & 43      \\ \hline
Anti-q$\bar{\rm q}$  &  $\acopl<179.$           & 1.69    & 26.1    & 38      \\
             &  $\ptsumf>0.035$         & 1.63    & 22.0    & 28      \\ \hline
Anti-WW      &  $\thrust>0.74$          & 1.33    & 16.3    & 23      \\
             &  $\plsumf<0.12$          & 1.25    & 12.1    & 19      \\ 
             &  $\ffntmi>0$             & 1.23    & 11.3    & 18      \\
             &  $\ffanmi>23.^\circ$     & 1.16    & 8.61    & 15      \\
             &  $(\ffanmi+\ffanma)>170.^\circ$ & 1.11    & 8.18    & 13      \\ \hline  
\end{tabular}
\end{tabular}
}
\label{tab:cut_class1}
\end{center}
\end{table}

\subsection{Class 2: Final state with more than one hard lepton}
\label{sec:class2}

\subsubsection{Class 2a: Two leptons and four jets\\{\it targeted channel:} $\mathrm{ZH \rightarrow \ell^+\ell^-\,q\bar{q}\,q\bar{q}}$}

The dominant backgrounds after preselection are $\mathrm{q\bar q}$, semi-leptonic WW as well as ZZ 
events where one of the Z bosons decays into hadrons and the other into leptons. 
The rejection of $\mathrm{q\bar q}$ and WW events is achieved by applying a cut on the variable $y_{45}$ and requiring the invariant mass of the two leptons (\MZrec) to be in a window around the nominal Z mass (\MZ). 
The remaining ZZ background is reduced by cutting on the angle between the two leptons and on their total energy.
The details of the preselection and selection are shown in Table~\ref{tab:cuts_2}.

The discriminant variable, inspired from the Higgs boson mass is then computed as:
\begin{equation}
D_1 = \sqrt{(\esum-\ELo-\ELt)^2-(\psum-\PLo-\PLt)^2}\ ,
\end{equation}
where \PLo{} and \PLt{} are the momenta of the leptons associated with the selected pair and 
$\psum$ is the total measured momentum. The discriminant variable is shown in Figs.~\ref{fig:2_final}a 
and~\ref{fig:2_final}b after preselection and final selection, respectively.
In this subclass, the targeted-signal efficiency is 74\%. The expected background is 0.67 events.

\subsubsection{Class 2t: Two leptons and missing energy\\{\it targeted channel:} $\mathrm{ZH \rightarrow \ell^+\ell^-\,\tau\nu\,q\bar{q}}$}
\label{sec:class2aT}

This is the only case where the one- or three-prong hadronic tau decays can be distinguished 
efficiently from other hadronic decays. 
The rejection of $\mathrm{q\bar q}$ and WW events is achieved by requiring the reconstructed 
mass of the two leptons to be in a window around the Z mass. A cut on the 
transverse momentum of the lepton pair ($P_t(\ell^+\ell^-)$) reduces the ZZ background. 
The selection cuts are shown in Table~\ref{tab:cuts_2}. In this subclass, the targeted-signal 
efficiency is 60\%. The expected background is 0.46 events.

\subsubsection{Class 2b: Two leptons, two jets and one soft lepton\\{\it targeted channel:} $\mathrm{ZH \rightarrow \ell^+\ell^-\,q\bar{q}\,\ell\nu}$}

This subclass is characterized by a third lepton and significant hadronic activity.
The selection proceeds in a similar way as in the subclass 2a by concentrating on the two 
leptons associated with the Z boson decay. The isolation for the lepton that is the most 
anti-parallel to the missing momentum ($I_{\ell_A}$) is used in the selection. The lepton 
tends to be more isolated in the signal than in the background. The selection 
cuts are shown in Table~\ref{tab:cuts_2}. After the final selection the dominant background is 
ZZ events decaying into $\mathrm{\ell^+\ell^-\bar b b}$. 

The pair of leptons which have the same flavour and opposite charge and give the best estimate of the Z mass are associated with the Z boson decay.
The following discriminant is used:
\begin{equation}
D_2 = \sqrt{(\esum-E_{\rm Z})^2-(\psum-\vec{P}_{\rm Z})^2+2(\ELth(\esum-E_{\rm Z})+\PLth(\psum-\vec{P}_{\rm Z}))}\ ,
\end{equation}
where $\vec{P}_{\rm Z}$ and $E_{\rm Z}$ are the momenta and energy of the Z boson, determined from the 
two assigned leptons. The additional term with respect to $D_1$ introduces 
the correction needed to take into account the undetected neutrino, assumed to be produced 
back-to-back to the third lepton.  The discriminant variable is shown in Figs.~\ref{fig:2_final}c and 
\ref{fig:2_final}d before and after the full selection, respectively.
In this subclass, the targeted-signal efficiency is 71\%.
The expected background is 0.16 events.

\subsubsection{Class 2c: Three leptons and two jets\\{\it targeted channel:} $\mathrm{ZH \rightarrow \ell^+\ell^-\,\ell\nu\,q\bar{q}}$}
\label{sec:class2c}

This subclass is characterized by a third lepton and low hadronic activity.
Compared to subclass 2b, the missing transverse momentum is higher, which makes 
it more difficult to distinguish from semi-leptonic WW events.
The mass window for the reconstructed Z mass is also broader due to the larger combinatorial background. 
A cut on the transverse momentum of the Z boson is applied to reduce the ZZ background.

\begin{figure}
\begin{center}
\begin{tabular}{cc}
\includegraphics[width=0.44\textwidth]{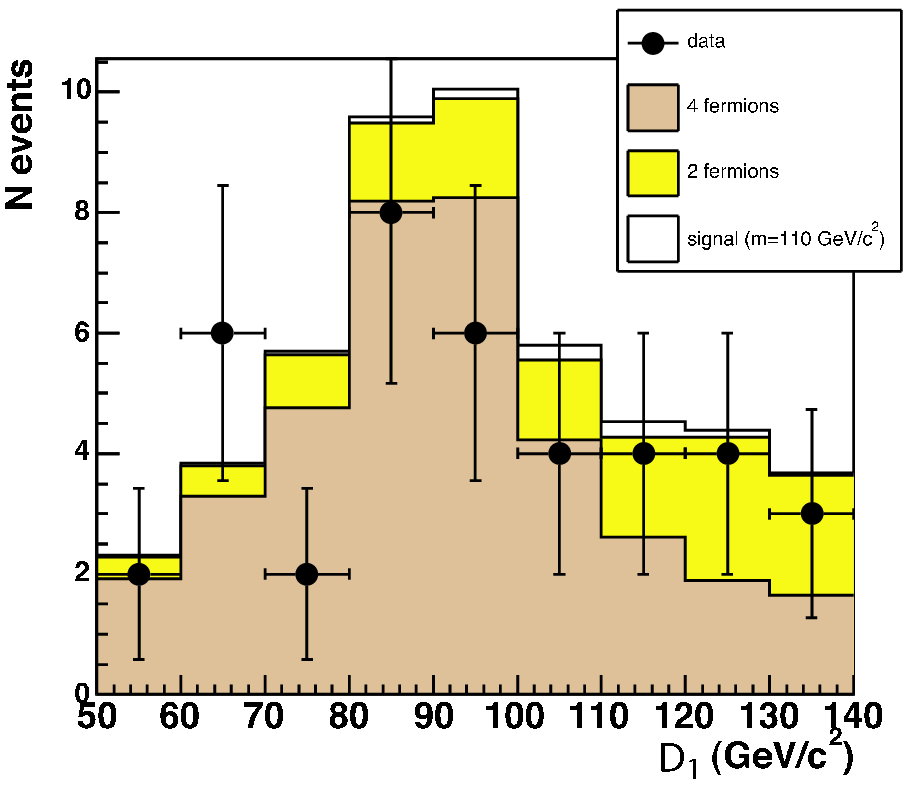} & \includegraphics[width=0.455\textwidth]{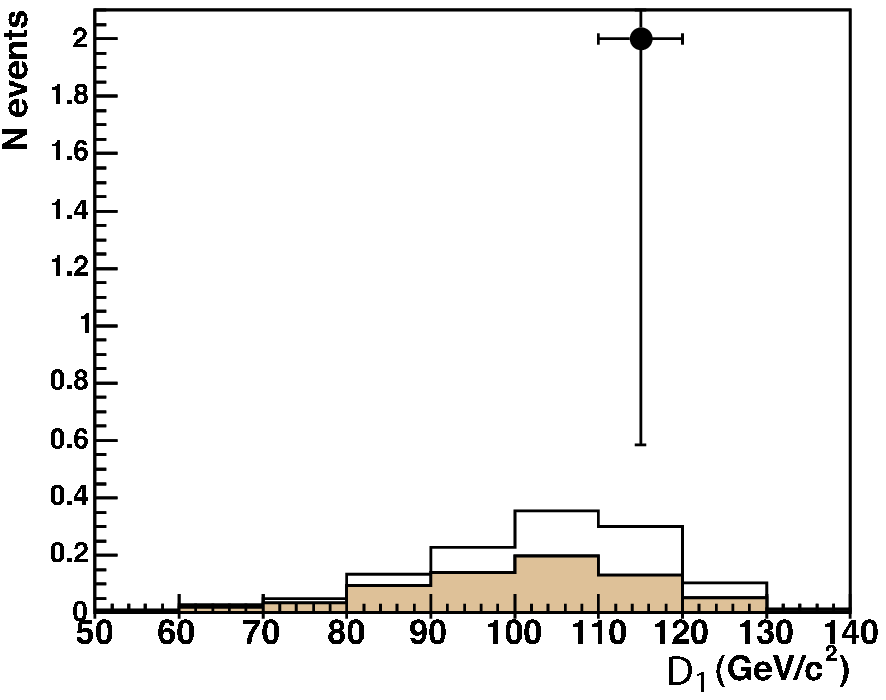}\\
\small (a) & \small (b)\\
\includegraphics[width=0.44\textwidth]{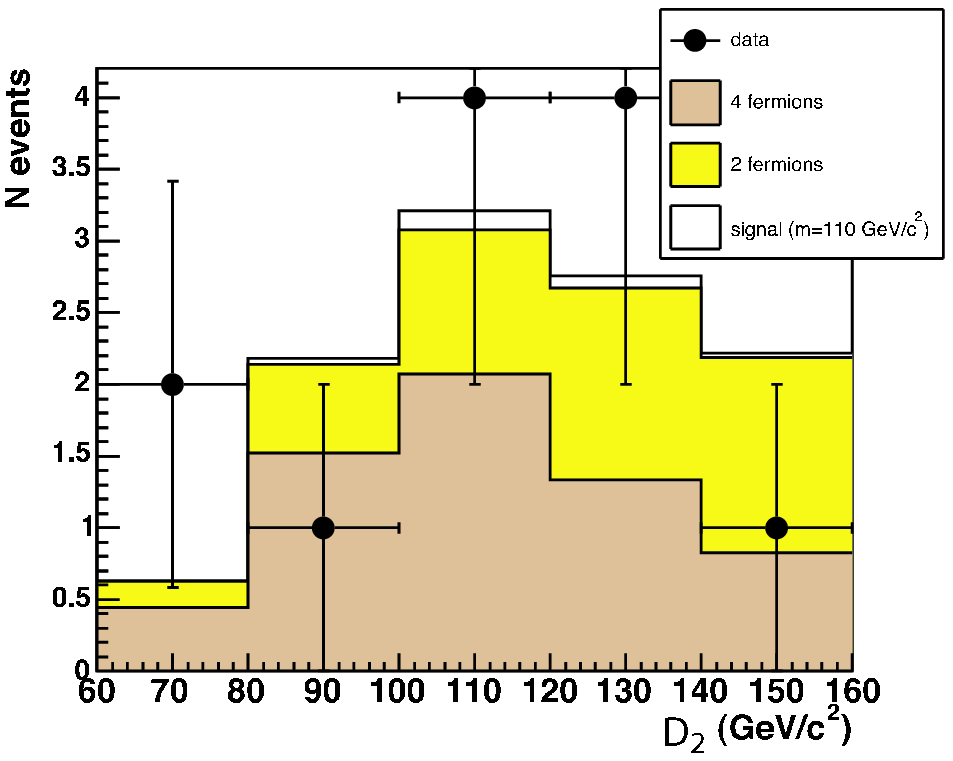} & \includegraphics[width=0.42\textwidth]{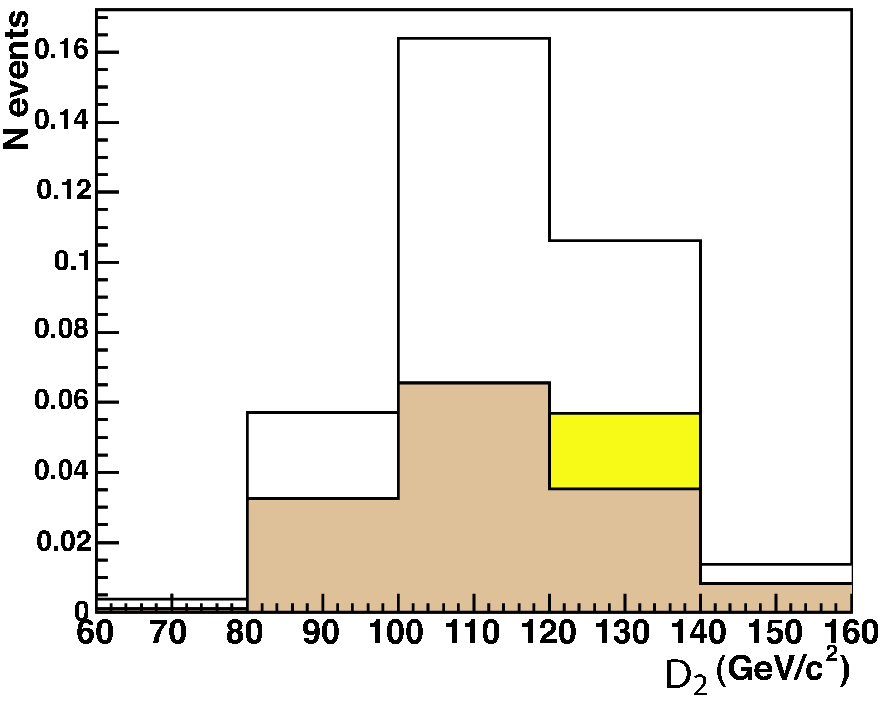}\\
\small (c) & \small (d)\\
\includegraphics[width=0.44\textwidth]{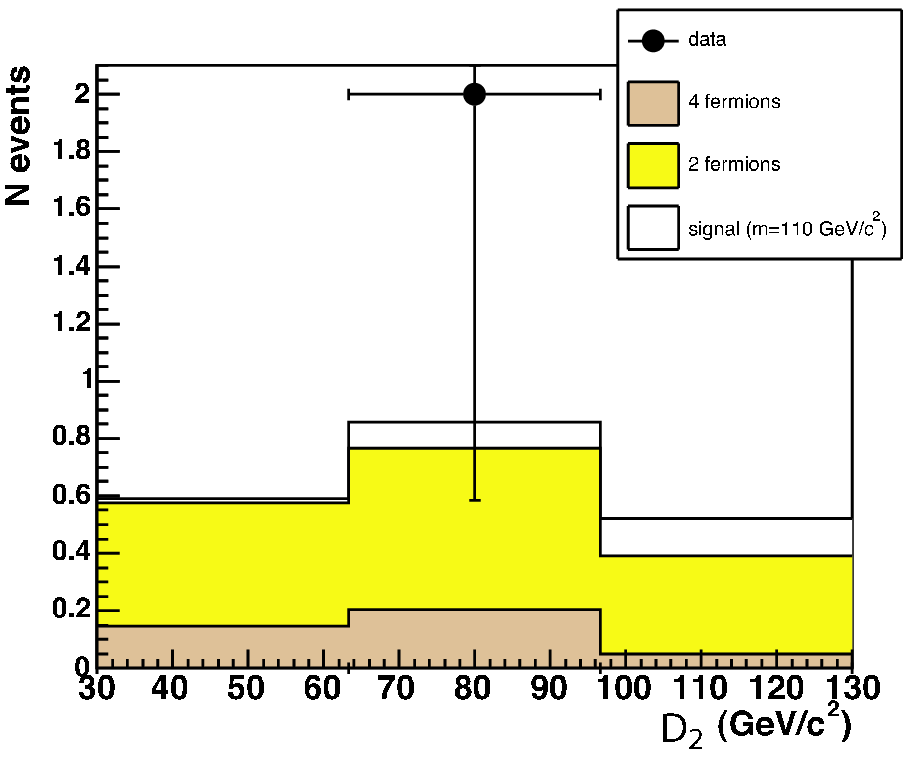} & \includegraphics[width=0.43\textwidth]{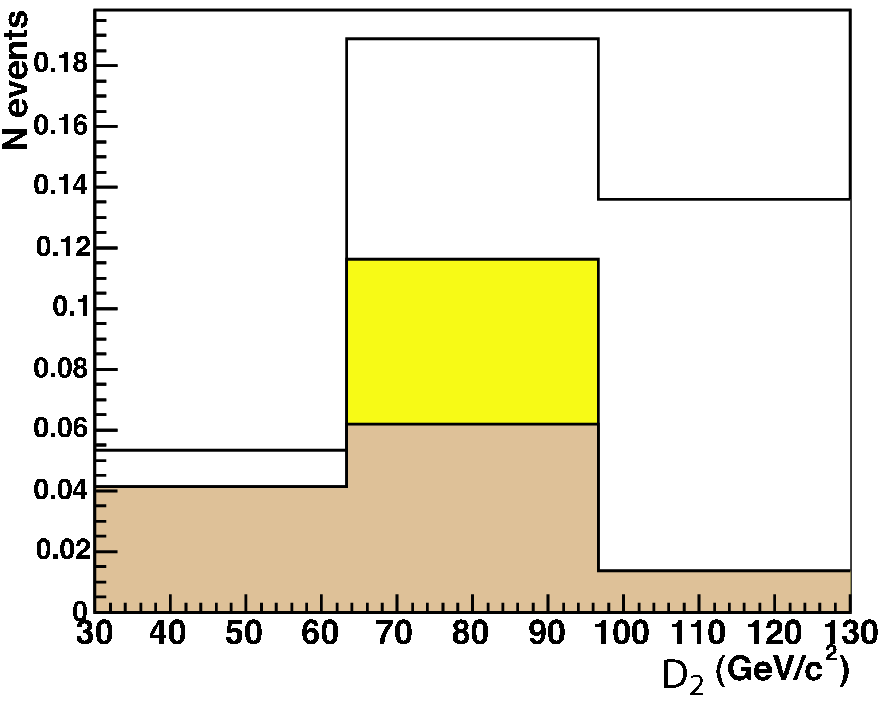}\\
\small (e) & \small (f) \\ 
\end{tabular}
\put(-95,270){\Large \bf ALEPH}
\caption{\small 
   Discriminant variable, $D_1$, for subclass 2a
   after the preselection (a) and the final selection (b). 
   In plots (c) and (d), the discriminant variable, 
   $D_2$, is shown after the preselection and the final selection, respectively,
   for events in class 2b.
   The two lower plots (e) and (f) show the same variable, 
   after the preselection and the final selection, respectively,
   for events in class 2c. All distributions are obtained from 2000 data.}
\label{fig:2_final}
\end{center}
\end{figure}

The signal and background distributions of the discriminant variable, $D_2$, are shown in 
Figs.~\ref{fig:2_final}e and~\ref{fig:2_final}f. The dominant background is  ZZ events.
In this subclass, the targeted-signal efficiency is 91\%. The expected background is 0.17 events.

\subsubsection{Class 2d: Three leptons plus one track\\{\it targeted channel:} $\mathrm{ZH \rightarrow \ell^+\ell^-\,\ell\nu\,\ell\nu}$}
\label{sec:class2d}

This subclass is characterized by a small branching fraction (the targeted channel represents 
$5\%$ of the events in class 2) but with a clear topology: four leptons in the final state, 
one of them being soft. To reduce the ZZ background, cuts on the thrust and the event acoplanarity are applied. 
The remaining WW events are rejected by requiring that the most anti-parallel 
lepton with respect to the missing momentum is well isolated. The selection criteria are 
detailed in Table~\ref{tab:cuts_2}. In this subclass, the targeted-signal efficiency is 57\%.
The expected background is 0.58 events.

\begin{table}
\begin{center}
\caption{\small
  Selection criteria for each subclass in class 2. 
  The numbers of signal ($N_{\mathrm s}$), background ($N_{\mathrm b}$) and data ($N_{\mathrm d}$) events
  are given in the table for the year 2000. 
  Energies, momenta  and masses are expressed in \gev, \gevc\ and \gevct, respectively.}
\vspace*{0.2cm}
\resizebox{\textwidth}{!}{
\hspace{-.3cm}
\begin{tabular}{ll}
\begin{tabular}{|p{3cm}|p{5.5cm}|c|c|c|}
\hline
 Subclass 2a   & Cuts                        & $N_{\mathrm s}$ & $N_{\mathrm b}$ & $N_{\mathrm d}$ \\ 
\hline
\hline
 Preselection  & \mbox{$\nch>8$}             & 1.03    & 48.8    & 39      \\
\hline
\hline
Anti-$\mathrm{q\bar q}$, WW 
               & \mbox{$\ln(\yffi)\geq-7.$}  &         &         &         \\
               & \mbox{$|\MZrec-\MZ|<14.$}   & 0.62    & 6.29    & 2       \\
\hline
Anti-ZZ        & \mbox{$\TLoLt>135.^\circ$}  &         &         &        \\
               & \mbox{$\ELo+\ELt < 95.$}    & 0.54    & 0.67    & 2      \\
\hline
\end{tabular}
& 
\begin{tabular}{|p{3cm}|p{5.5cm}|c|c|c|}
\hline
 Subclass 2t   & Cuts                        & $N_{\mathrm s}$ & $N_{\mathrm b}$ & $N_{\mathrm d}$ \\
\hline
\hline
 Preselection  & \mbox{$\ptsumf>0.002$}      & 0.08    & 2.59    & 2       \\
\hline 
\hline
Selection      & \mbox{$|\MZrec-\MZ|<23.$}             &         &         &         \\
               & \mbox{$P_t(\ell^+\ell^-) < 60.$}      & 0.06    &  0.46   & 1       \\
&&&&\\&&&&\\
\hline
\end{tabular}
\\
&\\
\begin{tabular}{|p{3cm}|p{5.5cm}|c|c|c|}
\hline
 Subclass 2b   & Cuts                        & $N_{\mathrm s}$ & $N_{\mathrm b}$ & $N_{\mathrm d}$  \\
\hline
\hline
 Preselection  & \mbox{$\nch>7$}             & 0.29    &  10.7   & 12       \\
\hline
\hline
Anti-$\mathrm{ q\bar q}$, WW 
               & \mbox{$\ln(\yffi)>-8.$}     &         &         &          \\
               & \mbox{$|\MZrec-\MZ|<20.$}   &         &         &          \\
               & \mbox{$\ln(I_{\ell_A})>-7.$}& 0.19    & 1.08    & 0        \\
\hline
  Anti-ZZ      & \mbox{$\TLoLt>142.^\circ$}  &         &         &           \\
               & \mbox{$\ELo+\ELt < 98.$}    &  0.18   & 0.16    & 0        \\
\hline
\end{tabular}
 & 
\begin{tabular}{|p{3cm}|p{5.5cm}|c|c|c|}
\hline
 Subclass 2c   & Cuts                        & $N_{\mathrm s}$ & $N_{\mathrm b}$ & $N_{\mathrm d}$  \\
\hline
\hline
 Preselection  & \mbox{$\ptsumf>0.01$}       &         &         &          \\ 
               & \mbox{$\nch>4$}             & 0.23    & 1.73    & 2        \\ 
\hline 
\hline
Anti-$\mathrm{q\bar q}$, WW 
               & \mbox{$\ln(\yffi)>-11.$}     &         &         &        \\
               & \mbox{$|\MZrec-\MZ|<23.$}    &         &         &          \\
               & \mbox{$\ln(I_{\ell_A})>-11.$}& 0.20    &  0.18   & 0       \\
\hline
  Anti-ZZ      & \mbox{$P_{t_{\rm Z}}<60.$}        & 0.20    & 0.17    & 0        \\
\hline
\end{tabular}
 \\
&\\
\begin{tabular}{|p{3cm}|p{5.5cm}|c|c|c|}
\hline
 Subclass 2d   & Cuts                        & $N_{\mathrm s}$ & $N_{\mathrm b}$ & $N_{\mathrm d}$  \\
\hline
\hline
 Preselection  & \mbox{$\ptsumf>0.11$}       & 0.10    & 3.13    & 3        \\
\hline 
\hline
 Anti-ZZ       & \mbox{$\thrust < 0.98$}     &         &         &          \\ 
               & \mbox{$\acopl < 176.^\circ$} & 0.09   &  0.93   & 1        \\
\hline
 Anti-WW       & \mbox{$\ln(I_{\ell_A})>-9.$} & 0.09   & 0.58    & 0        \\
\hline
\end{tabular}
 &
\end{tabular}
}
\label{tab:cuts_2}
\end{center}
\end{table}

\subsection{Class 3: Final state with one hard lepton}
\label{sec:class3}

In class 3, in order to reduce the $\gamma\gamma$ background, additional preselection cuts are
applied on the total transverse momentum and on the cosine of the polar angle of the 
missing momentum \cthmiss. Depending on the subclass, a cut on the number of 
tracks and/or on the most energetic reconstructed photon (\Ego) is required.

\subsubsection{Class 3a: One lepton and four jets\\{\it targeted channel:} $\mathrm{ZH \to q\bar{q}\,\ell\nu\,q\bar{q}}$}

To eliminate the $\mathrm{\gamma\gamma}$, $\mathrm{\ell^+\ell^-}$ and $\mathrm{q\bar q}$ backgrounds, 
cuts are applied to the track multiplicity, 
the angle between the hard lepton and the total momentum \tLPtot, and the mass of the hard W boson, 
reconstructed as the invariant mass of the hard lepton and the missing momentum (\MLPmiss).
To suppress the remaining backgrounds, mainly WW, \ELo, the hadronic acollinearity (\nlacolin) the thrust computed without the hard lepton (\nlthrust), $\yffi$, and the total transverse momentum are used.
The selection criteria together with the numbers of signal, background and data events are summarized in Table~\ref{tab:cuts_3}.

Figures.~\ref{fig:3ab_final}a and \ref{fig:3ab_final}b show the discriminant variable, \nlmsumn, after 
the preselection and after all cuts, respectively. In this subclass, the targeted-signal efficiency is 39\%.
The expected background is 2.9 events.

\subsubsection{Class 3b: One lepton, two jets plus one soft lepton\\{\it targeted channel:} $\mathrm{ZH \to q\bar{q}\,\ell\nu\,\ell\nu}$}

The selection procedure is very similar to that in subclass 3a, and is shown together with the numbers of signal, background and data events after each selection step in Table~\ref{tab:cuts_3}. 
The $\mathrm{\gamma\gamma}$, $\mathrm{\ell^+\ell^-}$ and $\mathrm{q\bar q}$ backgrounds are rejected by cuts on the number of tracks, on the total acollinearity and on the transverse momentum.
To further suppress $\mathrm{q \bar q}$ events, cuts on the angle between the hard lepton and the total momentum as well as on the event sphericity are applied.
Finally, most of the WW events are removed by a cut on the hard lepton energy which is sensibly smaller for signal events than for WW events, on the hadronic acollinearity, on $\nlyot$, computed without the leptons, and on $\ythf$.
Remaining background events are mainly semi-leptonic WW decays with a soft lepton produced in a jet. 

\begin{figure}
\begin{center}
\begin{tabular}{cc}
\includegraphics[width=0.44\textwidth]{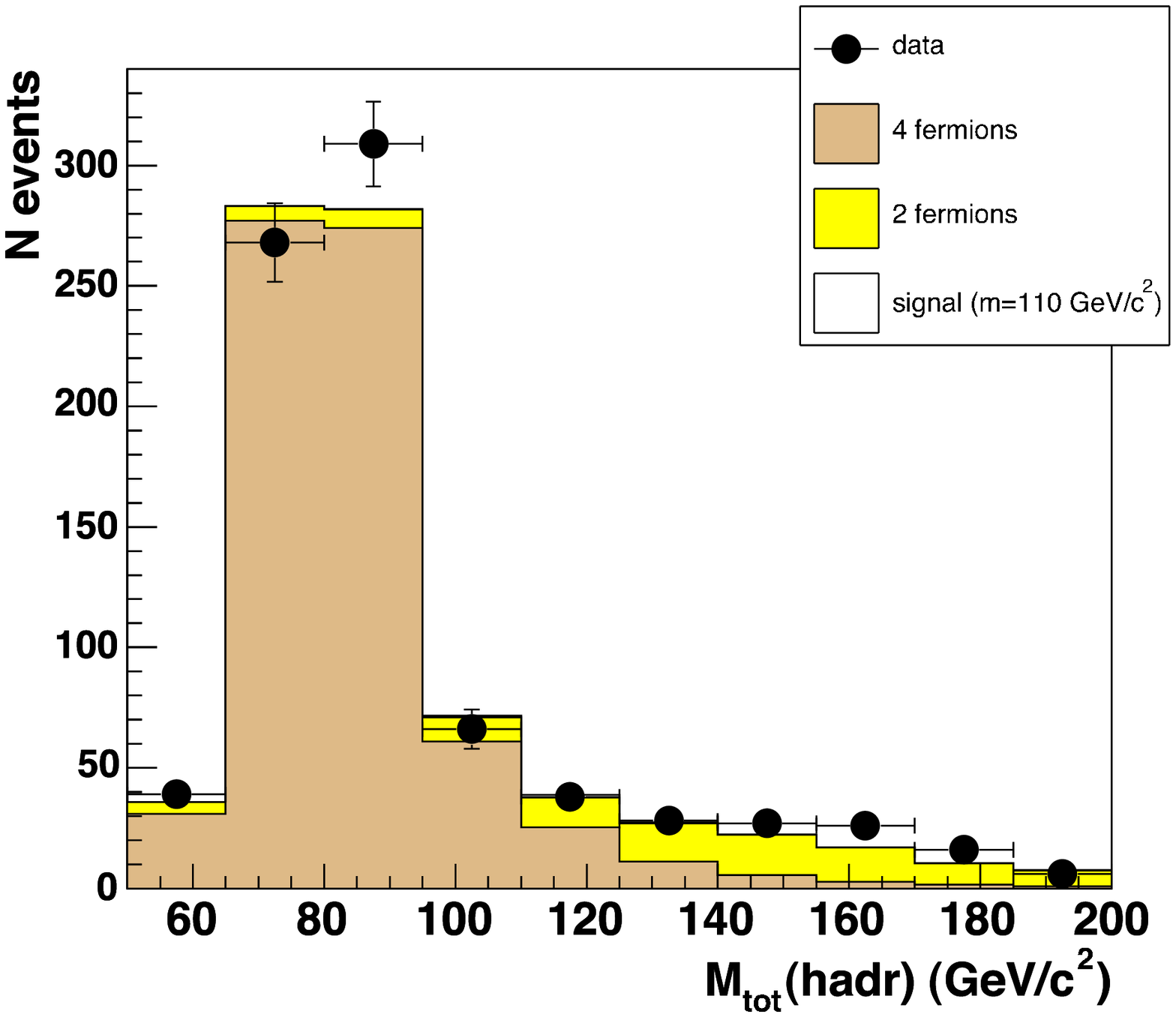} & \includegraphics[width=0.453\textwidth]{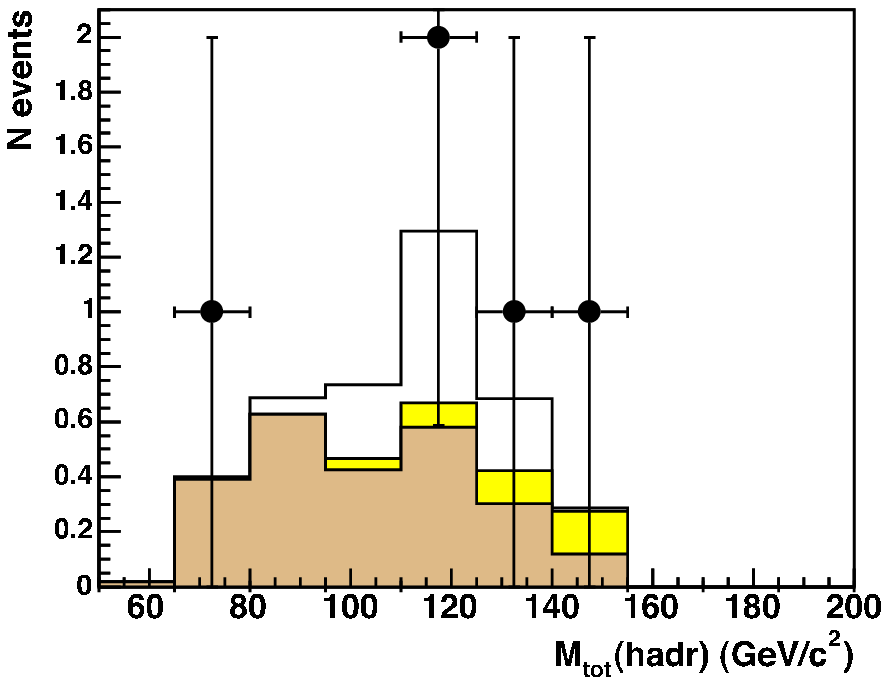}\\
\small(a) & \small(b) \\
\includegraphics[width=0.44\textwidth]{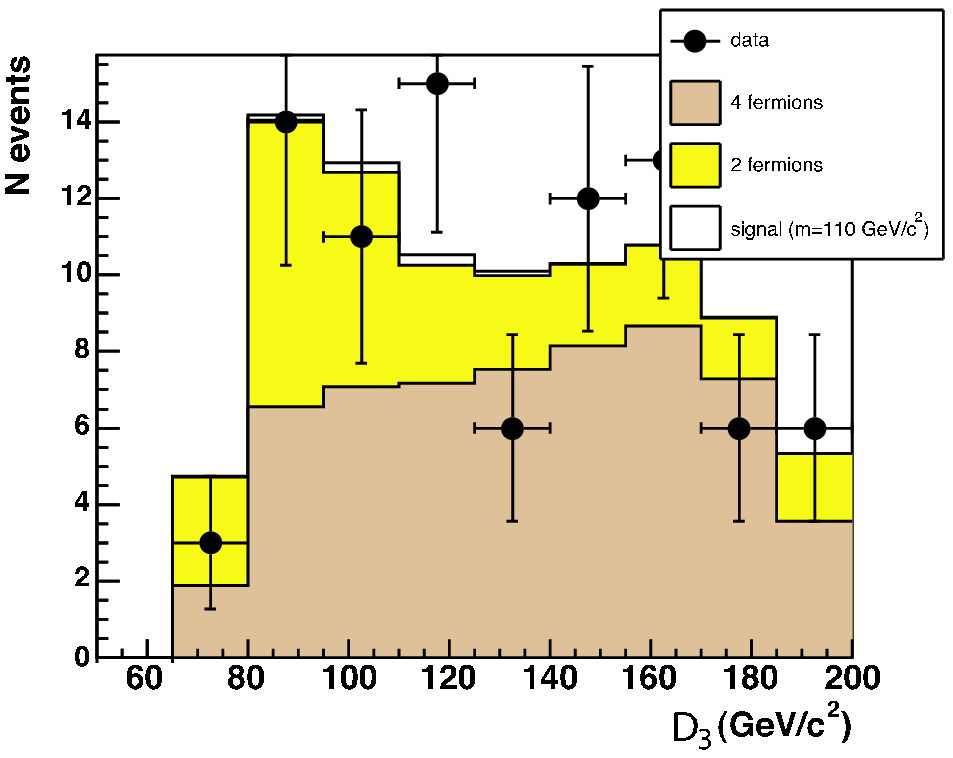} & \includegraphics[width=0.39\textwidth]{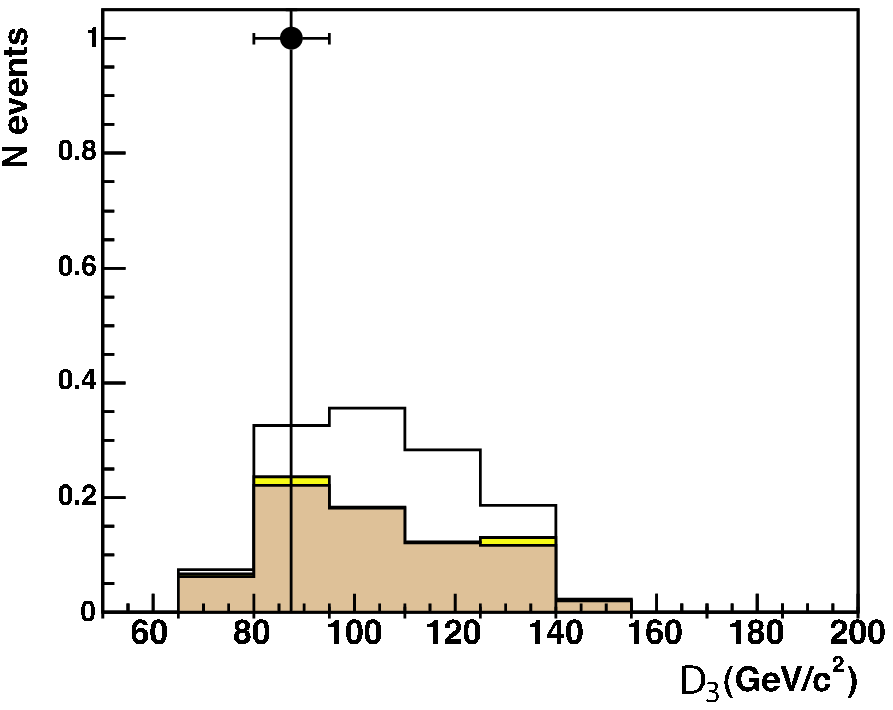}\\
\small(c) & \small(d) \\
\end{tabular}
\put(-95,175){\Large \bf ALEPH}
\caption{\small 
  Discriminant variable, \nlmsumn, for subclass 3a events after the preselection (a) and after the 
  final selection cuts (b).
  Discriminant variable, $D_3$, for subclass 3b events after the preselection (c) and after 
  the final selection cuts (d). All distributions are obtained from 2000 data.
}
\label{fig:3ab_final}
\end{center}
\end{figure}

A good estimate of the Higgs boson mass can be obtained from twice the sum of the lepton energies. The following variable is therefore used as discriminant variable:
\begin{equation}
D_3=2(\ELo+\ELt).
\label{eqn3}
\end{equation}
The discriminant variable, $D_3$, after the preselection and after all cuts 
is shown in Figs.~\ref{fig:3ab_final}c and~\ref{fig:3ab_final}d respectively. In this 
subclass, the targeted-signal efficiency is 60\%. The expected background is 0.76 events.

\subsubsection{Class 3c: One lepton and one track \\{\it targeted channel:} $\mathrm{ZH \to \nu\bar{\nu}\,\ell\nu\,\ell\nu}$}

This subclass is characterized by a significant acollinearity between the two leptons, and by a small 
invariant mass. Only events with exactly two tracks are kept. 
The $\mathrm{\gamma \gamma}$ events are rejected by a cut on the acollinearity and 
on the transverse momentum. 
The total mass and the acoplanarity are used to reject $\mathrm{\ell^+\ell^-}$ events 
as well as part of the WW background. A final cut on the total missing mass and on \tLPtot\ 
removes most of the remaining WW and ZZ background.

The discriminant variable, $D_3$, after the preselection and after all cuts is given in 
Figs.~\ref{fig:3cd_final}a and~\ref{fig:3cd_final}b respectively.
The selection criteria together with the numbers of signal, background and data events are 
summarized in Table~\ref{tab:cuts_3}.
In this subclass, the targeted-signal efficiency is 58\%.
The expected background is 3.5 events.

\subsubsection{Class 3d: One lepton and two jets \\{\it targeted channel:} $\mathrm{ZH \to \nu\bar{\nu}\,\ell\nu\,q\bar{q}}$}

Events in this subclass are characterized by a single hard lepton with some soft hadronic activity. The
$\mathrm{\gamma \gamma}$ and $\mathrm{\ell^+\ell^-}$ events are rejected by cutting on the 
aplanarity, \aplan~\cite{thrust}. The dominant WW background is then reduced in two stages.
First, the angle between the hard lepton and the total momentum and the energy of both the first and second most
energetic leptons are used. Then, a final rejection is achieved by cuts on the sphericity, on \nlmsumn\ and on \nlyot.

\begin{figure}
\begin{center}
\begin{tabular}{cc}
\includegraphics[width=0.44\textwidth]{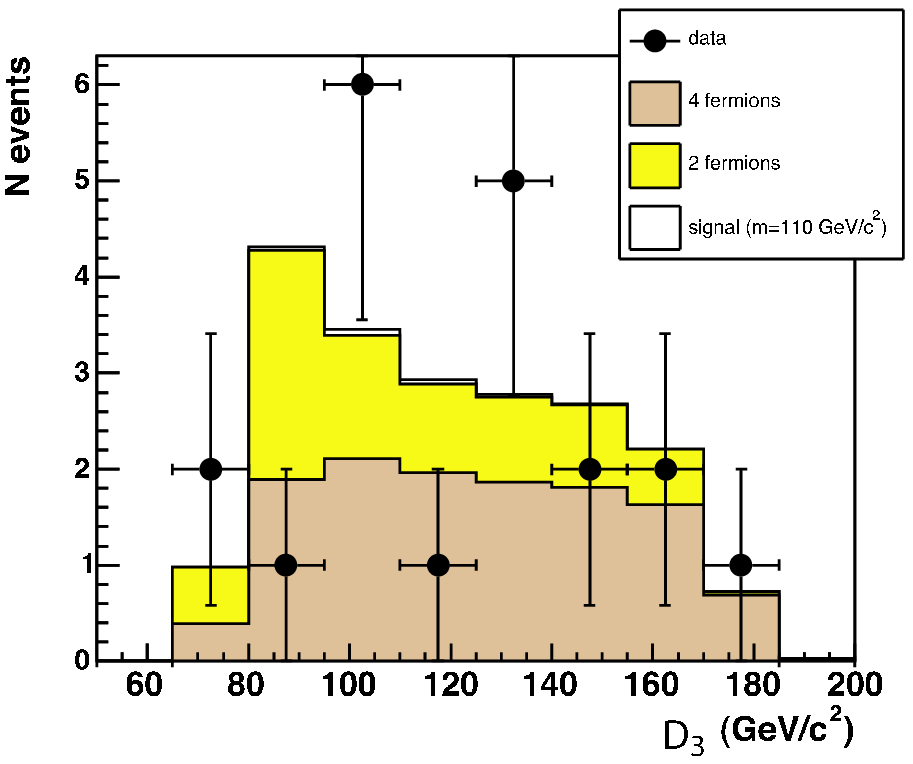} & \includegraphics[width=0.41\textwidth]{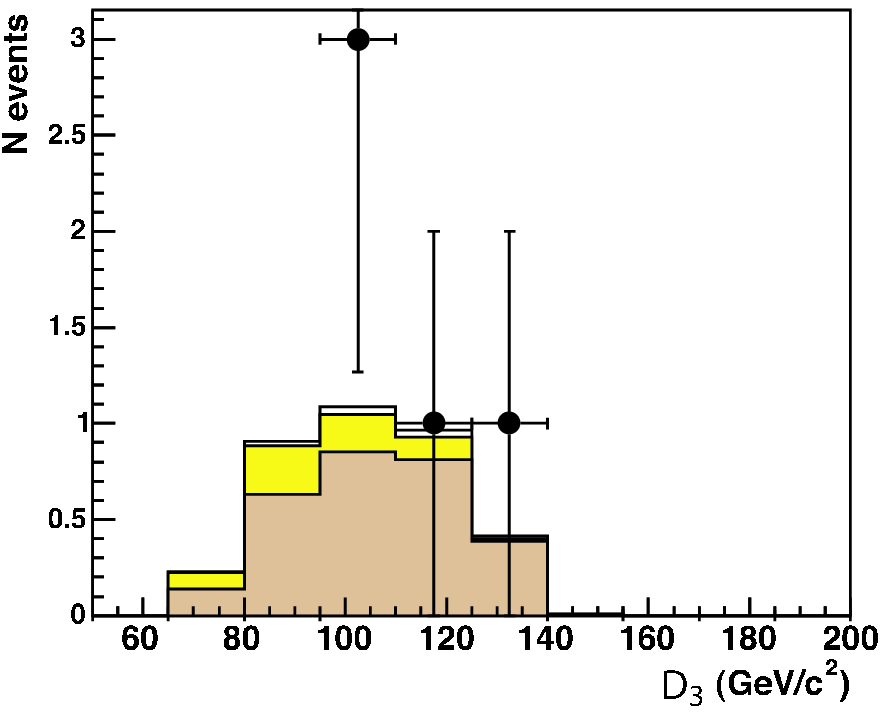}\\
\small (a) & \small (b) \\
\includegraphics[width=0.44\textwidth]{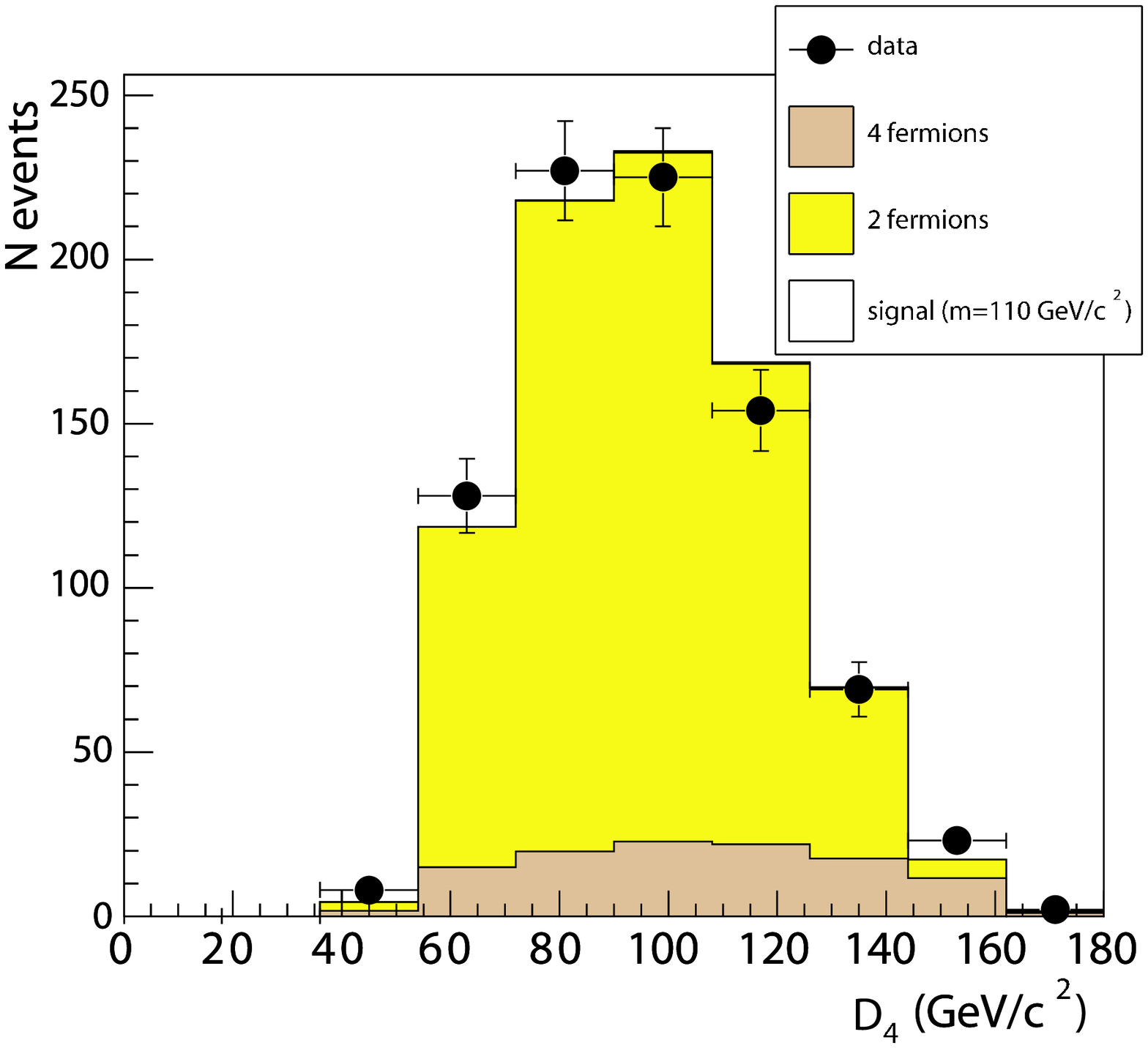} & \includegraphics[width=0.44\textwidth]{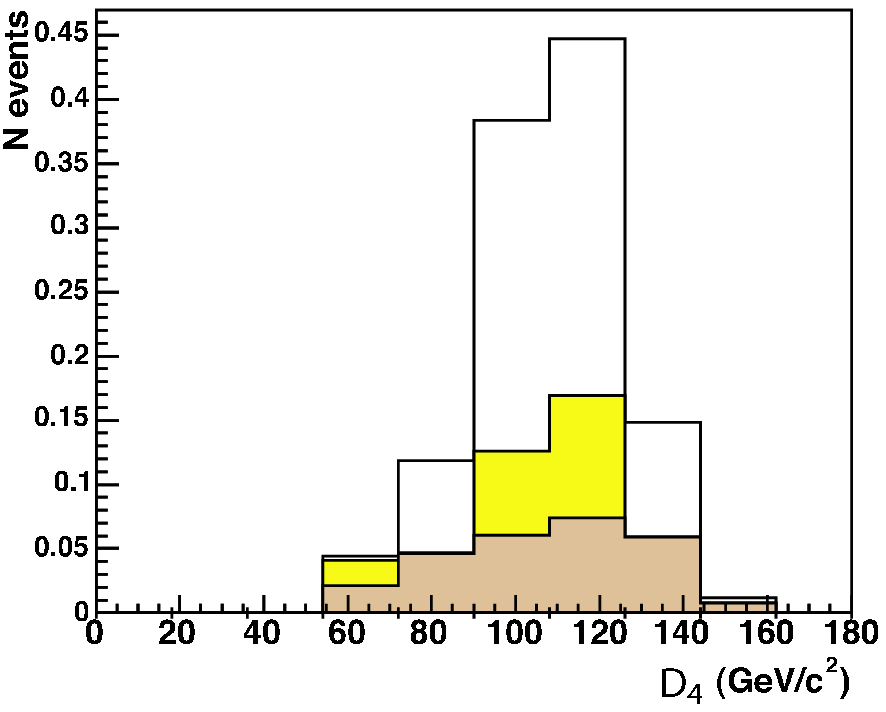} \\
\small (c) & \small (d) \\
\end{tabular}
\put(-95,185){\Large \bf ALEPH}
\caption{\small 
  Discriminant variable, $D_3$, for subclass 3c events after the preselection (a) and 
  the final selection (b). 
  Discriminant variable, $D_4$, for subclass 3d events after the preselection (c) and the final selection (d). 
  All distributions are obtained from 2000 data.
}
\label{fig:3cd_final}
\end{center}
\end{figure}

The hadronic activity is included in the discriminant variable, defined as
\begin{equation}
D_4=\sqrt{(\esum+\ELo)^2-(\psum-\PLo)^2}.
\label{eqn4}
\end{equation}
The discriminant variable after the preselection and after all cuts is shown in 
Figs.~\ref{fig:3cd_final}c and~\ref{fig:3cd_final}d respectively.
The selection criteria together with the numbers of signal, background and data events are 
summarized in Table~\ref{tab:cuts_3}. In this subclass, the targeted-signal efficiency is 58\%.
The expected background is 0.65 events.

\begin{table}
\begin{center}
\caption{\small 
  Selection criteria for each subclass in class 3. 
  The numbers of signal ($N_{\mathrm s}$), background ($N_{\mathrm b}$) and data ($N_{\mathrm d}$) events
  are given in the table for the year 2000. 
  Energies, momenta  and masses are expressed in \gev, \gevc~ and \gevct, respectively.}
\vspace*{0.2cm}
\resizebox{\textwidth}{!}{
\hspace{-.3cm}
\begin{tabular}{cc}
\begin{tabular}{|l|p{6cm}|c|c|c|}
\hline
 Subclass 3a    & Cuts                         & $N_{\mathrm s}$ & $N_{\mathrm b}$ & $N_{\mathrm d}$\\
\hline
\hline
 Preselection   & \mbox{$\ptsumf>0.05$}        &         &         &        \\
                & \mbox{$|\cthmiss|<0.9$}      &         &         &        \\
                & \mbox{$\nch>3$}              & 2.7     & 793.4   & 823   \\
                &                              &         &         &        \\
\hline
\hline
 Anti-q$\bar{\rm q}, \ell\ell, \gamma\gamma$  
                & \mbox{$\nch>20$}             &         &         &        \\
                & \mbox{$\tLPtot<41.^\circ$}   &         &         &        \\
                & \mbox{$\MLPmiss>55.$}        & 1.57    & 37      & 43     \\
\hline
 Anti-WW        & \mbox{$\ELo<55.$}            &         &         &        \\
                & \mbox{$\nlacolin>137.^\circ$}&         &         &        \\
                & \mbox{$\nlthrust<0.93$}      & 1.42    & 5.7     & 12     \\
\hline
 Anti-WW        & \mbox{$\ln(\yffi)>-7.2$}     &         &         &        \\
                & \mbox{$\ptsumf<0.25$}        & 1.24    & 2.9     & 5      \\
                &                              &         &         &        \\
\hline
\end{tabular}
&
\begin{tabular}{|l|p{6cm}|c|c|c|}
\hline
Subclass 3b     & Cuts                         & $N_{\mathrm s}$ & $N_{\mathrm b}$ & $N_{\mathrm d}$\\
\hline
\hline
 Preselections  & \mbox{$\ptsumf>0.05$}        &         &         &        \\
                & \mbox{$|\cthmiss|<0.9$}      &         &         &        \\
                & \mbox{$\nch>3$}              &         &         &        \\
                & \mbox{$\Ego<40.$}            & 0.85    & 86.96   & 86     \\
\hline
\hline
 Anti-q$\bar{\rm q}, \ell\ell, \gamma\gamma$    
                & \mbox{$\nch>5$}              &        &          &        \\
                & \mbox{$\acolin<178.^\circ$}  &        &          &       \\
                & \mbox{$\ptsumf>0.08$}        & 0.79   & 55.40    & 52     \\
\hline
 Anti-q$\bar{\rm q}$
                & \mbox{$\tLPtot<41.^\circ$}   &        &          &        \\
                & \mbox{$\ln(\spher)>-2.2$}    & 0.59   & 3.63     & 5      \\
\hline
 Anti-WW        & \mbox{$\ELo<55.$}            &        &          &        \\
                & \mbox{$\nlacolin>135.^\circ$}&        &          &        \\
                & \mbox{$\nlyot>0.23$}         &        &          &        \\
                & \mbox{$\ln(\ythf)>-6.$}      & 0.49   & 0.76     & 1      \\
\hline
\end{tabular}
 \\
&\\
\begin{tabular}{|l|p{6cm}|c|c|c|}
\hline
 Subclass 3c    & Cuts                         & $N_{\mathrm s}$ & $N_{\mathrm b}$ & $N_{\mathrm d}$\\
\hline
\hline
 Preselections  & \mbox{$\ptsumf>0.05$}        &         &         &        \\
                & \mbox{$|\cthmiss|<0.9$}      &         &         &        \\
                & \mbox{$\Ego<15.$}            & 0.17    & 19.9    & 20     \\
                &                              &         &         &        \\
\hline
\hline
 Anti-$\gamma\gamma$
                & \mbox{$\nch=2$}             &          &         &         \\
                & \mbox{$\acolin<138.^\circ$} &          &         &        \\
                & \mbox{$\ptsumf>0.09$}       & 0.15     & 11.16   & 14     \\
\hline
 Anti-$\ell\ell$, WW 
                & \mbox{$\msumf<0.26$}        &          &         &        \\
                & \mbox{$\acopl<160.^\circ$}  & 0.14     & 7.9     & 9      \\
\hline
 Anti-WW, ZZ    & \mbox{$\mmiss>127.$}        &          &         &        \\
                & \mbox{$\tLPtot<32.^\circ$}  & 0.12     & 3.5     & 5      \\
\hline
\end{tabular}
 &
\begin{tabular}{|l|p{6cm}|c|c|c|}
\hline
 Subclass 3d    & Cuts                        & $N_{\mathrm s}$ & $N_{\mathrm b}$ & $N_{\mathrm d}$ \\
\hline
\hline
 Preselections  & \mbox{$\ptsumf>0.05$}       &          &         &        \\
                & \mbox{$|\cthmiss|<0.9$}     &          &         &       \\
                & \mbox{$\nch>3$}             &          &         &        \\
                & \mbox{$\Ego<15.$}           & 1.22     & 829.8   & 836    \\
\hline
\hline
 Anti-$\gamma\gamma, \ell\ell$   
                & \mbox{$\ln(\aplan)>-8$}     & 0.94     & 4.92    & 3      \\
\hline
 Anti-WW        & \mbox{$\tLPtot<57.^\circ$}  &          &         &        \\
                & \mbox{$\ELo<60.$}           &          &         &        \\
                & \mbox{$\ELt<6.$}            & 0.90     & 3.39    & 1      \\
\hline
 Anti-WW        & \mbox{$-5.<\ln(\spher)<-2.$}&          &         &        \\
                & \mbox{$\nlmsum<0.17$}       &          &         &        \\
                & \mbox{$\ln(\nlyot)>-4.$}    & 0.79     & 0.65    & 0      \\
\hline
\end{tabular}
\end{tabular}
}
\label{tab:cuts_3}
\end{center}
\end{table}

\subsection{Class 4: Soft-lepton final state}
\label{sec:class4}

\subsubsection{Class 4a: One soft lepton and four jets\\{\it targeted channel:} $\mathrm{ZH \to q\bar{q}\,q\bar{q}\,\ell\nu}$}

The event selection relies on \yffi, \nlythf\ and the lepton isolation,
computed for the most anti-parallel lepton with respect to the missing momentum. 
This last variable reduces the background by a factor 6 (for a 42\% efficiency in signal events).
A $\chi^2$ is then built which takes into account the W and Z boson masses, reconstructed from the four jets in the event. 
For this, jets are paired and the mass of each pair is compared either to the nominal W boson mass or to the nominal Z boson mass.
The jet pairing that minimizes the $\chi^2$ is retained:
\begin{equation}
\chi^2={\frac {(M_{12}-m_{\rm Z})^2} {\sigma^2}}+{\frac {(M_{34}-m_{\rm W})^2}{\sigma^2}},
\label{eqnchi2cl4}
\end{equation}
were $M_{12}$ and $M_{34}$ are the masses of each jet pair and $\sigma$ is the estimated mean resolution on the reconstructed mass.
The background is finally reduced by constraining the value of the $\chi^2$, 
the total hadronic mass and the lepton energy.
The discriminating variable used for the estimation of the quoted confidence level is the reconstructed off-shell W mass, estimated from the missing momentum and the soft lepton momentum as follows:
\begin{equation}
D_{5}=\sqrt{(\emiss+\ELo)^2-(\pmiss+\PLo)^2}.
\label{eqn:cl4a}
\end{equation}  
Details of the event selection are given in Table~\ref{tab:cuts_4}.
The discriminant variable, after the preselection and the final 
selection is presented in Figs.~\ref{fig:4_final}a and \ref{fig:4_final}b, respectively.
In this subclass, the targeted-signal efficiency is 35\%.
The expected background is 4.0 events.

\subsubsection{Class 4b: One soft lepton and two jets\\{\it targeted channel:} $\mathrm{ZH \to \nu\bar{\nu}\,q\bar{q}\,\ell\nu}$}

Events are selected by cutting on \yot\ and \ythf, on the number of tracks and on the lepton energy.
The remaining background, still 75 times larger than the signal, is reduced by additional 
cuts on the total hadronic mass, the lepton isolation and on the hadronic acollinearity.
Since at least three neutrinos are expected in this channel, a cut on the missing mass is used in the final selection.
The dominant background after the selection is WW.

\begin{figure}
\begin{center}
\begin{tabular}{cc}
\includegraphics[width=0.44\textwidth]{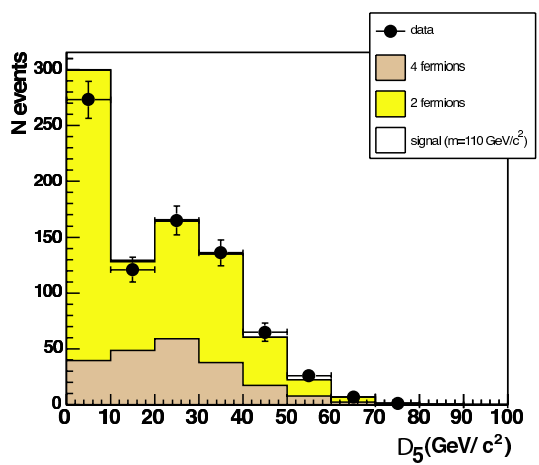} & \includegraphics[width=0.43\textwidth]{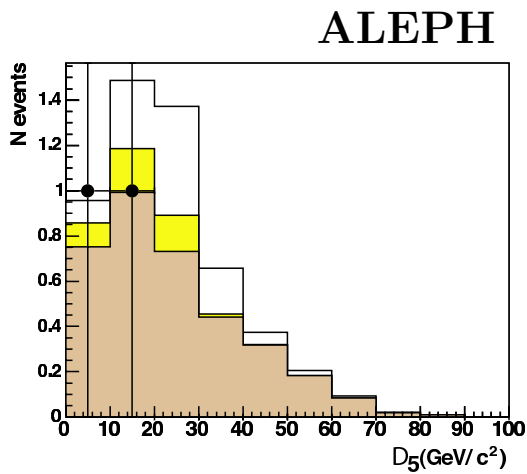}\\
\small (a) & \small (b) \\
\includegraphics[width=0.44\textwidth]{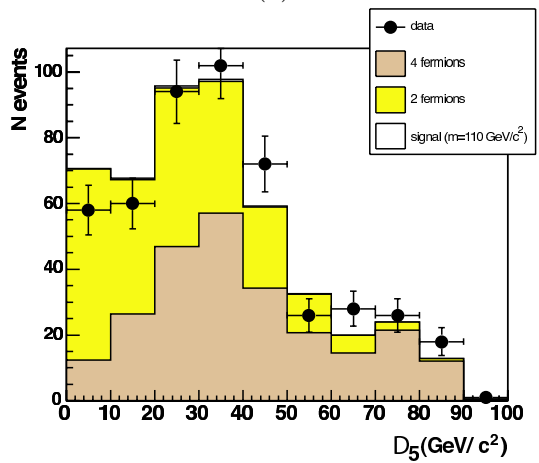} & \includegraphics[width=0.43\textwidth]{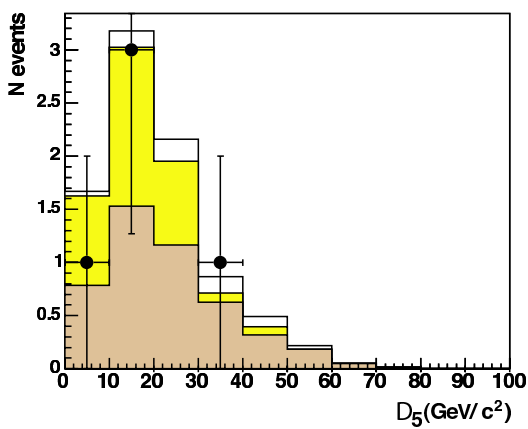}\\
\small (c) & \small (d) \\
\end{tabular}
\put(-85,185){\Large \bf ALEPH}
\caption{\small Discriminant variable for subclass 4a events after the preselection (a) and after the 
    final selection cuts (b). Discriminant variable for subclass 4b events after 
    the preselection (c) and after the final selection cuts (d). All distributions are obtained from 2000 data.}
\label{fig:4_final}
\end{center}
\end{figure}

The discriminant variable, $D_5$, after the preselection and the final selection is presented in Figs.~\ref{fig:4_final}c and \ref{fig:4_final}d, respectively.
The expected number of events after the signal selection is shown in Table~\ref{tab:cuts_4}. 
In this subclass, the targeted-signal efficiency is 54\%.
The expected background is 8.0 events.

\begin{table}
\begin{center}
\caption{\small  
    Selection criteria for each subclass in class 4.
    The numbers of signal ($N_{\mathrm s}$), background ($N_{\mathrm b}$) 
    and data ($N_{\mathrm d}$) events are given in the table for the year 2000. 
    Energies, momenta and masses are expressed in \gev, \gevc~ and \gevct, 
    respectively.}
\vspace*{0.2cm}
\resizebox{\textwidth}{!}{
\begin{tabular}{cc}
\begin{tabular}{|l|p{6cm}|c|c|c|}
\hline
 Subclass 4a    & Cuts                        & $N_{\mathrm s}$ & $N_{\mathrm b}$ & $N_{\mathrm d}$ \\
\hline
\hline
 Preselections  & \mbox{$\nch>10$}            &         &         &         \\
                & \mbox{$|\cthmiss|<0.95$}    & 4.6     & 818     & 794     \\
                &                             &         &         &         \\
\hline
\hline
 Group 1        & \mbox{$\ln(\yffi)>-6.$}     &         &         &         \\
                & \mbox{$\ln(\nlythf)>-4.6$}  &         &         &         \\ 
                & \mbox{$\ln(I_{\ell_A})> -3.5 $}& 1.4     & 7.4     & 8       \\
                &                             &         &         &         \\
\hline
 Group 2        & \mbox{$\chi^2 < 0.02$}      &         &         &         \\
                & \mbox{$\nlmsum<0.95$ }      &         &         &         \\
                & \mbox{$E_{\ell_A}>7. $}        & 1.17    & 4.0     & 2       \\
                &                             &         &         &         \\
\hline
\end{tabular}
 & 
\begin{tabular}{|l|p{6cm}|c|c|c|}
\hline
 Subclass 4b    & Cuts                        & $N_{\mathrm s}$ & $N_{\mathrm b}$ & $N_{\mathrm d}$ \\
\hline
\hline
 Preselections  &  \mbox{$\nch>3$}            &         &         &         \\
                &  \mbox{$\ptsumf>0.05$}      &         &         &         \\
                &  \mbox{$|\cthmiss|<0.95$}   & 2.4     & 478     & 485     \\
\hline
\hline
 Group 1        & \mbox{$\ln(\nlyot)>-7.$}    &         &         &         \\
                & \mbox{$\ln(\nlytth)>-2.6$}  &         &         &         \\
                & \mbox{$\nch>9$}             &         &         &         \\
                & \mbox{$E_{\ell_A}>7.$}         & 1.35    & 99.7    & 82      \\
\hline
 Group 2        & \mbox{$\nlmsumn \in [50.,103.]$}&     &         &         \\
                & \mbox{$\ln(I_{\ell_A})>-4.5$}  &         &         &         \\
                & \mbox{$\nlacolin>140.^\circ$}&        &         &         \\
                & \mbox{$\mmiss>65.$}         & 0.69    & 8.0     & 5       \\
\hline
\end{tabular}
\end{tabular}
}
\label{tab:cuts_4}
\end{center}
\end{table}

\section{Results}
\label{sec:res}

The numbers of signal ($m_{\rm H}=110\gevct$), background and data events for the years 1999 and 2000,
for each class, are given in Table~\ref{tab:summary}.
The numbers of observed events agree well with the expectations for all subclasses but 3a and 3c, 
where there is a small excess ($\sim 2\sigma$) over the expected background (\clb=0.97).
For subclass 3a, 6.6 background events remain after the cuts, for 1.27 signal events, 
while thirteen candidates are observed. Similarly, for subclass 3c, 0.15 signal events 
are expected for 6.6 background events, and fourteen candidates are observed in the data.
The best sensitivity is achieved for subclasses with a low expected \ecls. 
These are subclasses 3d, 3a, 4a, 2a and 1b, in that order.

The four classes are combined and the compatibility between data
and background with and without the signal is evaluated with the
log-likelihood ratio estimator $\mathrm{ln\,Q}$~\cite{junk}.
The combined expected values of the signal and background confidence levels are 0.08 and 0.50 respectively.
The observed values for the signal and background confidence levels are 0.26 and 0.87, respectively.

\begin{table}[t]
\begin{center}
\caption{\small Numbers of signal events for a 110\gevct\ Higgs boson ($N_{\rm s}$), 
                background ($N_{\rm b}$) and data ($N_{\rm d}$) events, 
                as well as the value of the expected and observed
                confidence levels for each subclass, for the years 1999 and 2000 together.}
\vskip 2ex
\begin{tabular}{|r@{ : }l||c|c|c|c|c|}
\hline
\multicolumn{2}{|l||}{Class}            & $N_{\rm s}$         & $N_{\rm b}$        & $N_{\rm d}$ & \ecls    & \cls    \\ \hline \hline
1a & $\mathrm{ZH\rightarrow q\bar{q}\,q\bar{q}\,q\bar{q}}$     & $6.80\pm0.06$   & $372.7\pm1.7$  & 360     & 0.60     & 0.50   \\
1b & $\mathrm{ZH\rightarrow\nu\bar{\nu}\,q\bar{q}\,q\bar{q}}$    & $1.35\pm0.03$   & $18.0\pm0.4$   & 20      & 0.58     & 0.79   \\ \hline
\multicolumn{2}{|c||}{Class 1 combined} & $8.15\pm0.07$   & $390.7\pm1.7$  & 380     & 0.44     & 0.53    \\ \hline
2a & $\mathrm{ZH\rightarrow \ell^+\ell^-\,q\bar{q}\,q\bar{q}}$     & $0.64\pm0.02$   & $2.41\pm0.07$  & 5       & 0.57     & 0.91    \\
2t & $\mathrm{ZH\rightarrow \ell^+\ell^-\,\tau\nu\,q\bar{q}}$ & $0.070\pm0.006$ & $1.21\pm0.08$  & 1       & 0.94     & 0.96    \\
2b & $\mathrm{ZH\rightarrow \ell^+\ell^-\,q\bar{q}\,\ell\nu}$     & $0.21\pm0.01$   & $0.57\pm0.04$  & 2       & 0.81     & 0.83    \\
2c & $\mathrm{ZH\rightarrow \ell^+\ell^-\,\ell\nu  \,q\bar{q}}$     & $0.24\pm0.01$   & $0.31\pm0.05$  & 0       & 0.79     & 0.79    \\
2d & $\mathrm{ZH\rightarrow \ell^+\ell^-\,\ell\nu\,\ell\nu}$     & $0.113\pm0.008$ & $1.33\pm0.12$  & 1       & 0.91     & 0.90    \\ \hline
\multicolumn{2}{|c||}{Class 2 combined} & $1.27\pm0.03$   & $5.82\pm0.17$  & 9       & 0.43     & 0.73    \\ \hline
3a & $\mathrm{ZH \to q\bar{q}\,\ell\nu\,q\bar{q}}$            & $1.27\pm0.03$   & $6.6\pm0.2$    & 13      & 0.47     & 0.84    \\
3b & $\mathrm{ZH \to q\bar{q}\,\ell\nu\,\ell\nu}$            & $0.58\pm0.02$   & $2.01\pm0.12$  & 3       & 0.62     & 0.66    \\
3c & $\mathrm{ZH \to \nu\bar{\nu}\,\ell\nu\,\ell\nu}$          & $0.150\pm0.009$ & $6.6\pm0.2$    & 14      & 0.95     & 0.99    \\
3d & $\mathrm{ZH \to \nu\bar{\nu}\,\ell\nu\,q\bar{q}}$          & $0.89\pm0.02$   & $0.99\pm0.15$  & 1       & 0.44     & 0.43    \\ \hline
\multicolumn{2}{|c||}{Class 3 combined} & $2.89\pm0.04$   & $16.2\pm0.4$   & 32      & 0.23     & 0.47    \\ \hline
4a & $\mathrm{ZH\rightarrow q\bar{q}\,q\bar{q}\,\ell\nu}$      & $1.41\pm0.03$   & $8.92\pm0.2$    & 8       & 0.55     & 0.40    \\
4b & $\mathrm{ZH\rightarrow \nu\bar{\nu}\,q\bar{q}\,\ell\nu}$    & $0.85\pm0.02$   & $20.34\pm0.4$   & 19      & 0.78     & 0.68    \\ \hline
\multicolumn{2}{|c||}{Class 4 combined} & $2.27\pm0.03$   & $29.3\pm0.5$   & 27      & 0.50     & 0.30    \\ \hline \hline
\multicolumn{2}{|c||}{All     combined} & $14.6\pm0.09$   & $441.9\pm1.9$  & 448     &  0.08    & 0.26    \\ \hline
\end{tabular}
\label{tab:summary}
\end{center}
\end{table}

\subsection{Systematic uncertainties}
\label{sec:syst}
The main contribution to the systematic error comes from the simulated 
statistics. The uncertainties from the production cross sections, from the simulation 
of the calorimeter energy scale and from hadronization processes are also taken into
account.
The beam background is conservatively taken into account in the analysis (via event selection variable $\ef$).
Uncertainties from the simulated statistics are included when combining the searches in the 
several subclasses by varying the expected signal and background levels bin per bin in a number of toy 
Monte Carlo experiments, while other uncertainties are taken into account in a correlated way 
by varying the expected signal and background levels coherently in the same direction 
for all subclasses and all bins. Each source of uncertainty is nevertheless handled independently, 
which corresponds to adding up their effect in quadrature.

For the four different classes, after applying the selection cuts, the main remaining background 
comes from WW pair production. The related uncertainty, associated to W decays into three jets, 
is taken to be 2\%.
This is a conservative estimate from the uncertainty on the $\alpha_s$ measurement~\cite{PDG}.
The systematic errors related to the simulation of the absolute energy scale of the calorimeters are determined using hadronic Z events and are found to be $\pm 0.9\%$ and $\pm 2\%$ for the electromagnetic and hadronic calorimeters, respectively. 
The effect of a possible miscalibration of the calorimeters is evaluated on simulated samples by scaling the electromagnetic and hadronic part of the measured energy independently by these amounts. The largest of the observed shifts for each calorimeter is combined in quadrature.
This leads to a mean uncertainty of the order of 3\% on the remaining background level.
The uncertainties originating from the hadronization model are evaluated 
by comparing WW events hadronized using the string model ({\tt JETSET~7.4}~\cite{jetset} Monte Carlo)
and with the colour dipole model ({\tt ARIADNE~4.10}~\cite{ariadne} Monte Carlo).
The associated mean systematic error is 6\%.
Both the hadronization and calorimetric uncertainties are evaluated separately for each
subclass. The uncertainties taken into account are presented in Table~\ref{tab:syst}.

\begin{table}[b]
\begin{center}
\caption{\small Levels of systematic uncertainties considered for each subclass. 
The corresponding total systematic uncertainty is also given in the last column, 
excluding the impact of the limited simulated statistics.} 
\begin{tabular}{|c|c|c|c|c|} 
\multicolumn{5}{c}{} \\
\hline
  \raisebox{-2.2ex}{Subclass} & \multicolumn{4}{|c|}{\raisebox{-0.5ex}{Systematic uncertainty sources}} \\
\cline{2-5}
 \rule{0pt}{4.6mm}           & Calorimetry & Hadronization & $W\to \mathrm{3\ jets} $ & Total\\
\hline 
1a & 0\% & 2\% & 2\% & 2.8\%\\
1b & 6\% & 8\% & 2\% & 10.2\%\\
2  & 0\% & 0\% & 2\% & 2\%\\
3a & 2\% & 4\% & 2\% & 4.9\%\\
3b & 0\% &14\% & 2\% & 14.1\%\\
3c & 2\% & 8\% & 2\% & 8.5\%\\
3d &12\% & 8\% & 2\% & 14.6\%\\
4a & 2\% & 0\% & 2\% & 2.8\%\\
4b & 0\% & 9\% & 2\% & 9.2\%\\
\hline
\end{tabular}
\label{tab:syst}
\end{center}
\end{table}

\subsection{Combined Likelihood Ratio}
\label{sec:likelihood}

\begin{figure}
\begin{center}
 \includegraphics[width=10cm]{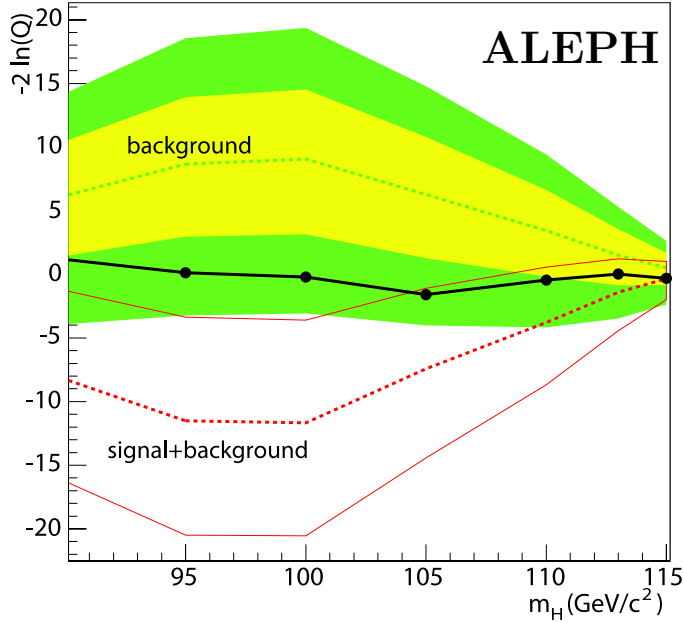}
 \put(-100,215){\Large \bf ALEPH}
 \caption{\small Log-likelihood ratio, $\mathrm{-2 ln\,Q}$, as a function 
of the Higgs boson mass hypothesis, $m_{\rm H}$, with all data collected between 
191\gev\ and 209\gev. 
The solid line is the result obtained from the data. 
The expected background-only and signal-plus-background likelihoods 
are indicated by the dashed lines; 
the light and dark shaded bands around the background expectation 
contain 68\% and 95\% of the simulated background-only experiments, 
respectively. The $1\sigma$ bands for the signal-plus-background 
hypothesis are also shown.}
 \label{fig:m2lnq_mass}
\end{center}
\end{figure}

The combined log-likelihood ratio as a function of the Higgs boson mass
is presented in Fig.~\ref{fig:m2lnq_mass}. A $1.5\sigma$ excess is observed. 
This excess does not depend on the hypothetical mass, 
and no supporting indication of a signal is seen.
An upper limit on the signal cross section is set as a function of the Higgs boson mass.

\subsection{Cross section upper limits}
\label{sec:upper}

\begin{figure}
\begin{center}
\includegraphics[width=7cm]{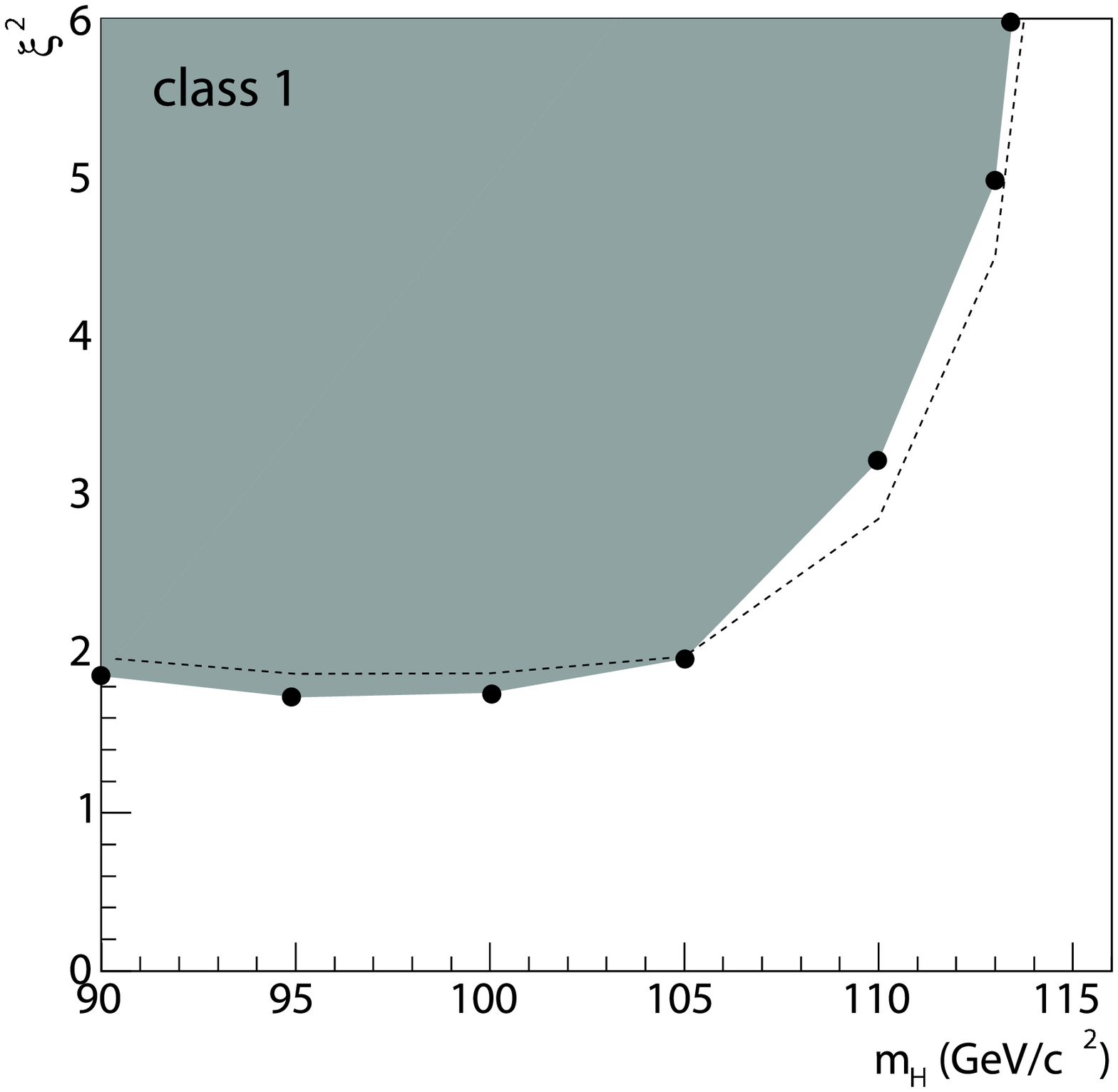}
\includegraphics[width=7cm]{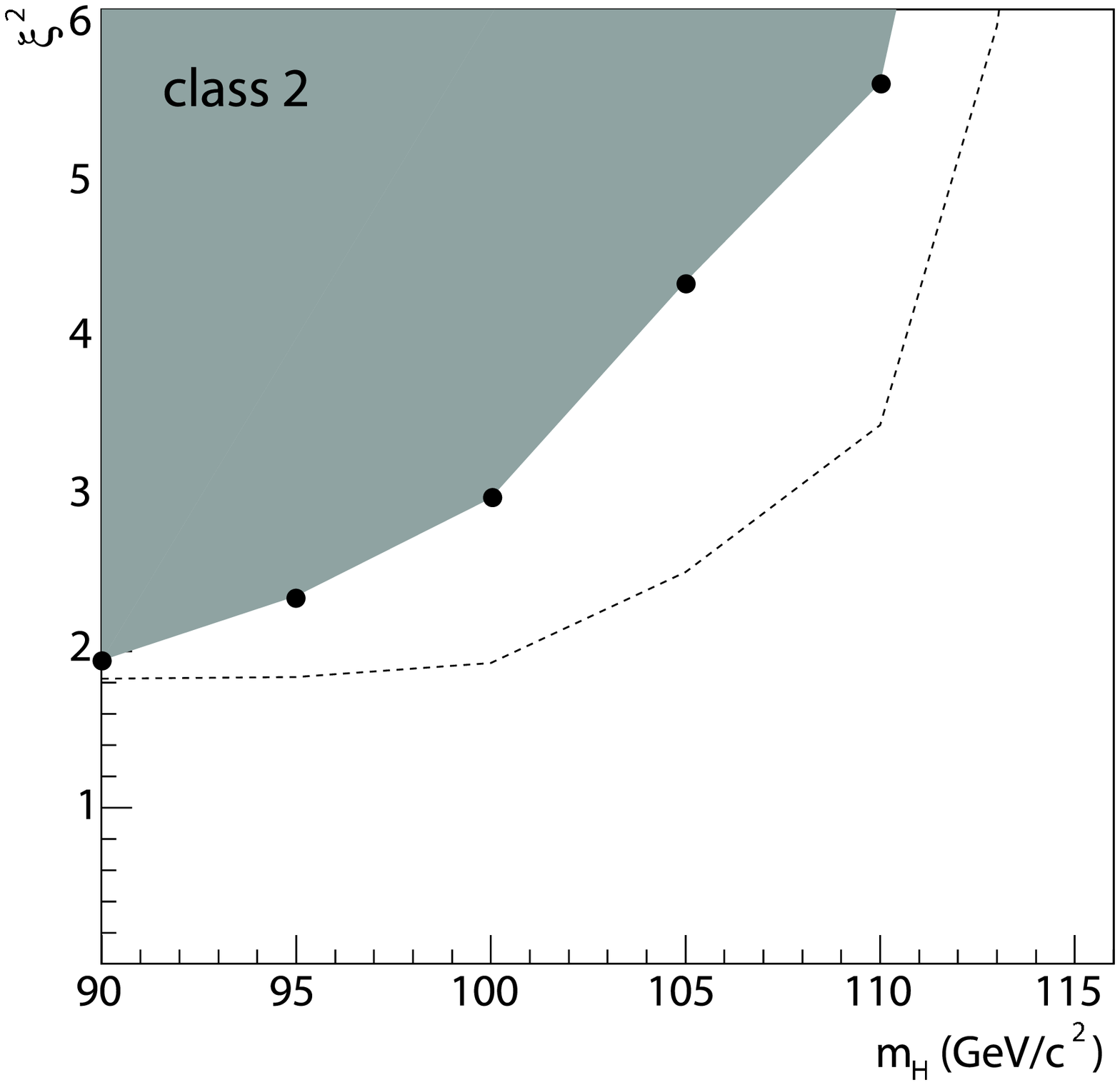}
\put(-95,180){\Large \bf ALEPH}
\\
\vskip 1mm
\includegraphics[width=7cm]{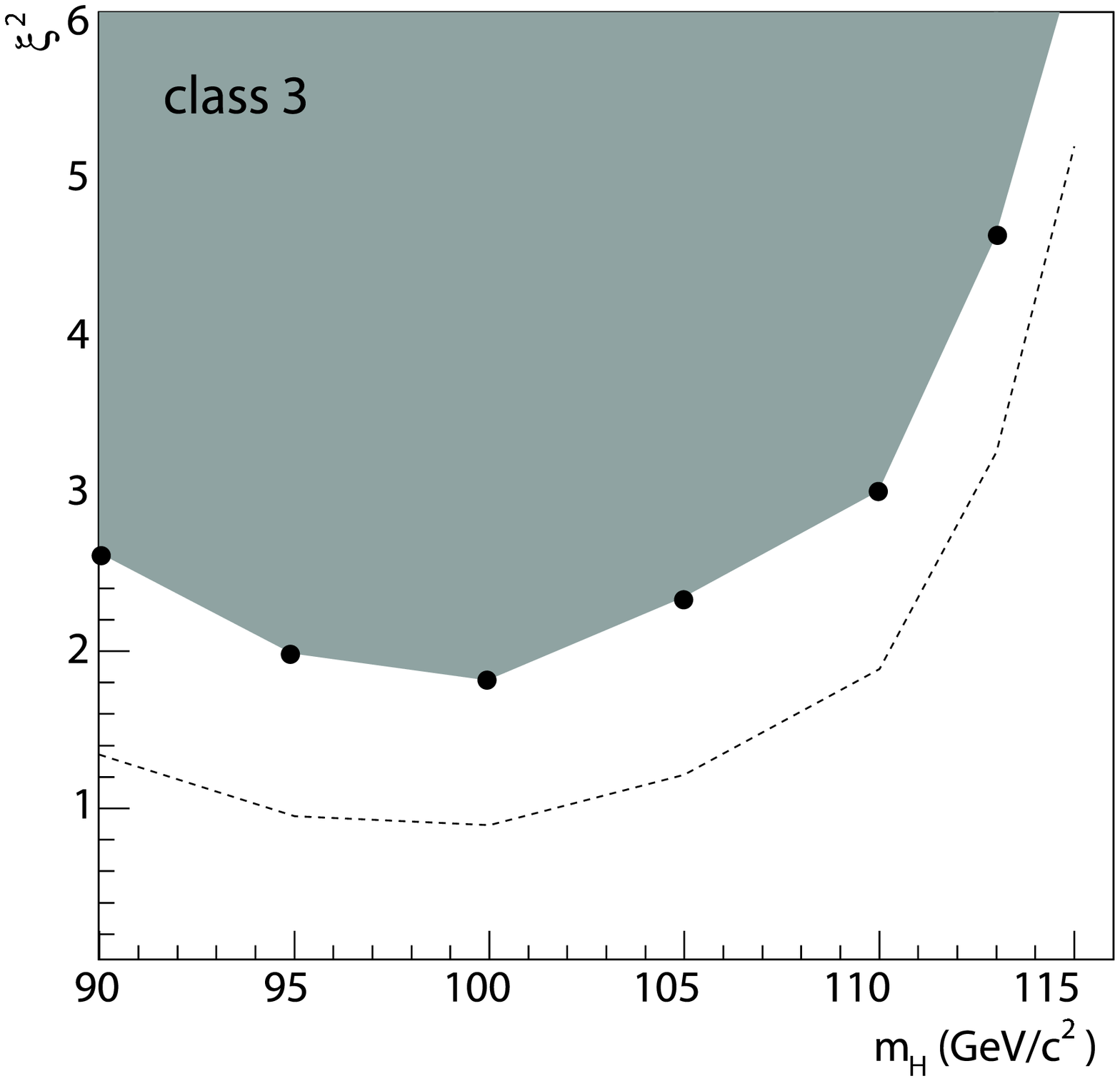}
\includegraphics[width=7cm]{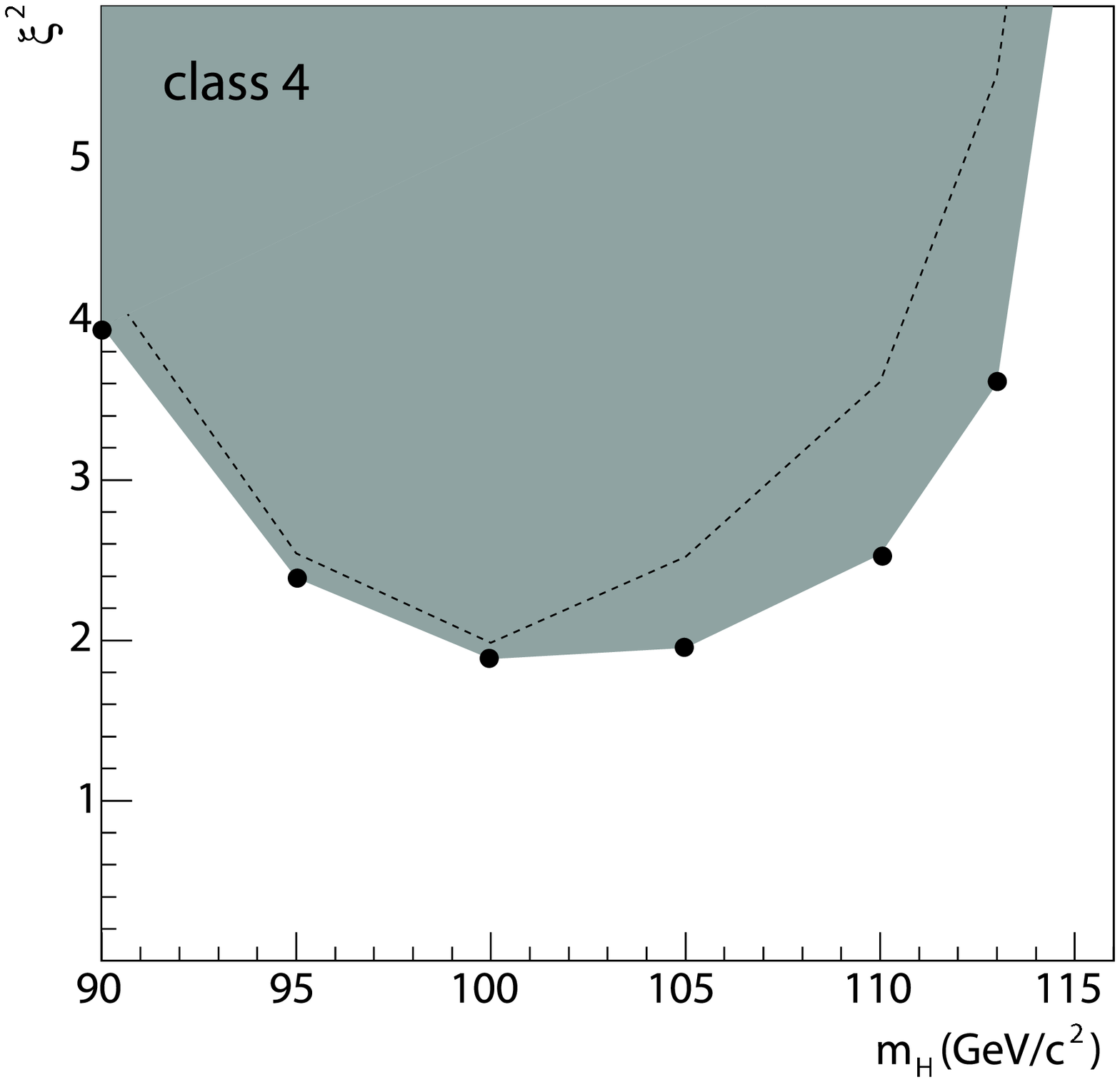}
 \caption{\small Limit on $\xi^2$ (defined in the text) as a function 
of the Higgs boson mass hypothesis, $m_{\rm H}$, in the four different classes. The dashed line 
corresponds to the expected limit while the 95\% C.L. excluded region 
is shown by the gray area.}
 \label{fig:sub_clplot}
\end{center}
\end{figure}

\begin{figure}[p]
\begin{center}
 \includegraphics[width=9.5cm]{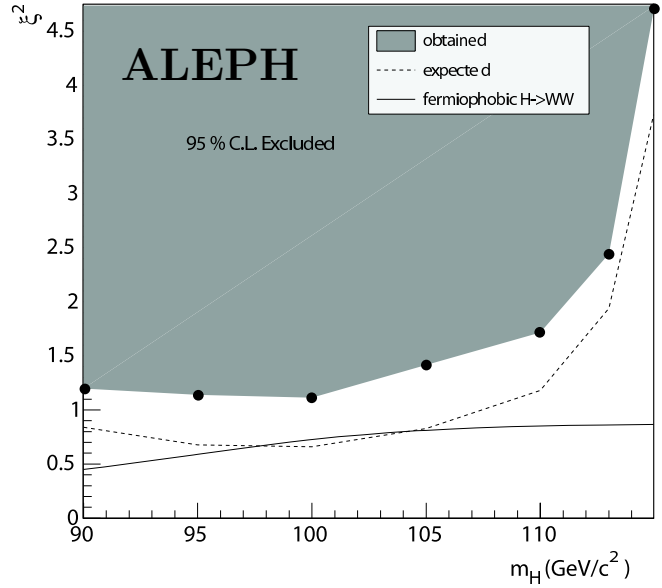}
\put(-225,200){\Large \bf ALEPH}
 \caption{\small Limit on $\xi^2$ (defined in the text) as a function 
of the Higgs boson mass hypothesis, $m_{\rm H}$. The dashed line corresponds to the expected 
limit while the 95\% C.L. excluded region is shown by the gray area. 
The benchmark fermiophobic Higgs model branching ratio is depicted by the full line.}
 \label{fig:clplot}
\end{center}
\end{figure}

For each class in the analysis, an upper limit on the production cross section 
at a given Higgs boson mass is derived. Figure~\ref{fig:sub_clplot} 
presents the resulting 95\% C.L. upper limit on 
$\mathrm{\xi^2 = BR(H\to WW) \sigma(e^+e^- \to Hf\bar{f}) / \sigma^{SM}(e^+e^- \to Hf\bar{f})}$, 
as a function of the Higgs boson mass. The third class, providing the best 
compromise between a clean signature and signal sensitivity 
 has the highest reach. The other three classes are nevertheless 
as important in order to get a competitive limit.
Combining all four analyses, the 95\% C.L. upper limit on $\xi^2$ 
as a function of the Higgs boson mass is given in Fig.~\ref{fig:clplot}. 
The benchmark fermiophobic Higgs model is drawn as the full line in 
the figure. The afore-mentioned excess in class 3 
results in an observed limit which is less than the expected limit.

\subsection{Mass exclusion limits}
\label{sec:massl}

Lower limits on the Higgs boson mass can be extracted in the context of a given model from upper 
limits on the cross section presented in the previous section.
If the branching ratio of the Higgs boson to W bosons is taken to be 100\%, the expected limit 
is 107.5\gevct. In the fermiophobic Higgs boson scenario, it is expected that Higgs boson masses between 97.5\gevct\ and 104\gevct\ can be excluded. 
For these two scenarios, due to the excess observed, no limit at 95\% C.L. can be set. 

However, the present search for $\mathrm{H \to WW}$ can be combined with the previously published ALEPH search for $\mathrm{H \to \gamma\gamma}$~\cite{Aleph_hgg} to significantly improve the limits on the fermiophobic Higgs boson scenario (Fig.~\ref{fig:branching}). 
The combined expected limit on the fermiophobic Higgs boson mass is 111.4\gevct, an improvement of $\sim7\gevct$ on the sensitivity from the $\mathrm{H \to \gamma\gamma}$ search alone~\cite{Aleph_hgg}\footnote{The actual expected limit quoted in Reference~\cite{Aleph_hgg}, 105.4\gevct, results from using a definition of \cls~\cite{jinMcnamara} which is different from the definition adopted in the present paper.}. The observed limit is 105.8\gevct.

\begin{figure}[p]
\begin{center}
 \includegraphics[width=9.5cm]{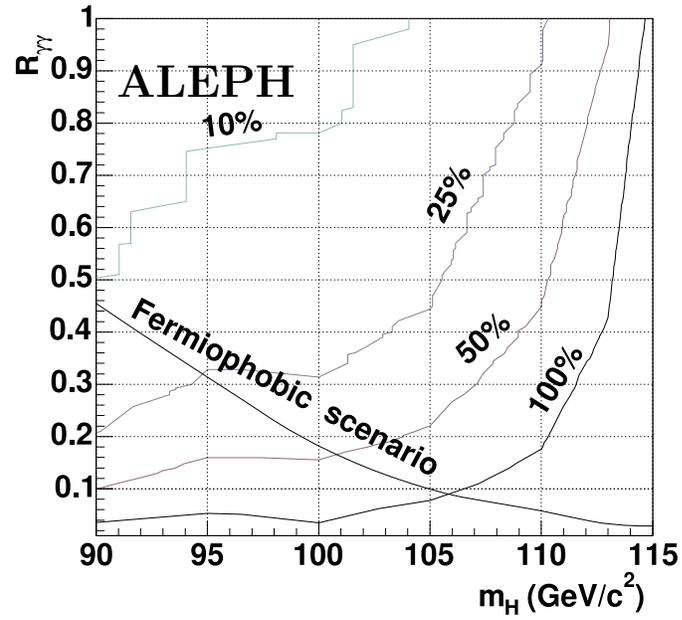}
\put(-220,200){\Large \bf ALEPH}
 \caption{\small The 95\% C.L. limit for $\mathrm{BR_{bosons}}$ as a function of $m_{\rm H}$ and $\mathrm{R_{\gamma\gamma}}$. The solid lines indicate the upper limit of exclusion regions. The crossing point between the ``$\mathrm{BR_{bosons}}=100\%$'' line and the ``Fermiophobic scenario'' line provides the lower limit on the Higgs boson mass in the benchmark scenario: $m_{\rm H} > 105.8\gevct$.}
 \label{fig:WWGGcombined}
\end{center}
\end{figure}

A model-independent limit can be derived by scanning the $\mathrm{H \to \gamma\gamma}$ and $\mathrm{H \to WW}$ branching fractions. This is conveniently parametrized as:
\begin{eqnarray*}
\mathrm{BR_{bosons}} & = & \mathrm{BR_{H\to\gamma\gamma}+BR_{H\to WW}+BR_{H\to ZZ}}, \\
\mathrm{R_{\gamma\gamma}} & = & \mathrm{BR_{H\to\gamma\gamma} / BR_{bosons}},
\end{eqnarray*}
where $\mathrm{R_{\gamma\gamma}}$ represents the fraction of bosonic decays into photon pairs 
and ranges from zero to one. 
The best limit is obtained combining the present results with those previously 
published by ALEPH on the search for $\mathrm{H \to \gamma\gamma}$~\cite{Aleph_hgg}.
The 95\% C.L. limit on $\mathrm{BR_{bosons} }$ is determined at each point of the $m_{\rm H}$ 
versus $\mathrm{R_{\gamma\gamma}}$ plane, resulting in the exclusion curves of 
Fig.~\ref{fig:WWGGcombined}.

\section{Conclusions}
\label{sec:conc}

A search for a Higgs boson produced in association with a Z and 
decaying into a WW pair has been performed with a dataset 
of $\mathrm{453.2\ pb^{-1}}$ recorded by the ALEPH detector
at centre-of-mass energies from 191 to 209\gev.
No statistically significant evidence for a fermiophobic Higgs boson decaying into a 
WW pair has been found in the data. 
Assuming Standard Model couplings to gauge bosons, a 95\% C.L. upper limit 
on the ratio $\mathrm{BR(H\to WW) \sigma(e^+e^- \to Hf\bar{f})} / \mathrm{\sigma^{SM}(e^+e^- \to Hf\bar{f})}$ 
has been obtained. 
Combining this analysis with the study of $\gamma\gamma$ decays of a 
Higgs boson~\cite{Aleph_hgg}, a Higgs boson mass up to 105.8\gevct\ 
has been excluded in the context of the benchmark fermiophobic scenario.
A model-independent limit has been derived 
by scanning the $\mathrm{H \to \gamma\gamma}$ and $\mathrm{H \to WW}$ branching fractions.
This analysis complements existing searches for new physics beyond the Standard 
Model and constrains models introducing anomalous couplings in 
the Higgs sector.

\clearpage

\section*{Acknowledgments}

It is a pleasure to congratulate our colleagues from the
accelerator divisions for the outstanding operation of 
LEP\,2, especially in its last year of running during which 
the accelerator performance was pushed beyond expectation. 
We are indebted to the engineers and technicians in all our 
institutions for their contributions to the excellent 
performance of the ALEPH detector. 
Those of us from non-member states 
wish to thank CERN for its hospitality and support.

\end{document}